\documentclass[]{aastex62}
\usepackage{tikz}
\usepackage{threeparttable}
\usepackage{amsmath}
\usepackage{lineno}
\usepackage{appendix}
\accepted{June 28, 2020}
\submitjournal{ApJ}
\newcommand{\boldred}[1]{#1}

\begin{document}
%\linenumbers
\title{Mapping circumstellar magnetic fields of late-type evolved stars with the Goldreich-Kylafis effect: CARMA observations at $\lambda 1.3$ mm of R Crt and R Leo}
\author{Ko-Yun Huang}
\affiliation{Department of Astronomy, University of Illinois at Urbana-Champaign, 1002 W. Green Street, Urbana, IL 61801 MC-221}
\author{Athol J. Kemball}
\affiliation{Department of Astronomy, University of Illinois at Urbana-Champaign, 1002 W. Green Street, Urbana, IL 61801 MC-221}
\author{Wouter H.T. Vlemmings}
\affiliation{Department of Space, Earth and Environment, Chalmers University of Technology, Onsala Space Observatory, 439 92 Onsala, Sweden}
\author{Shih-Ping Lai}
\affiliation{Institute of Astronomy and Department of Physics, National Tsing Hua University, Hsinchu 30013, Taiwa}
\author{Louis Yang}
\affiliation{Kavli Institute for the Physics and Mathematics of the Universe (IPMU), University of Tokyo, Japan}
\author{Iv\'an Agudo}
\affiliation{Instituto de Astrofisica de Andalucia (CSIC), Glorieta de la Astronomia s/n 18008 Granada, Spain}
%CARMA?
%1.3 mm spectral-line polarization observation with CARMA towards AGB stars}

\keywords{Asymptotic giant branch stars -- stellar magnetic fields -- polarimetry -- CO line emission -- circumstellar masers}

\section*{Abstract}
Mapping magnetic fields is the key to resolving what remains an unclear physical picture of circumstellar magnetic fields in late-type evolved stars. 
Observations of linearly polarized emission from thermal molecular line transitions due to the Goldreich-Kylafis (G-K) effect provides valuable insight into the magnetic field geometry in these sources that is complementary to other key studies.
In this paper, we present the detection of spectral-line polarization from both the thermal $J=2-1$ CO line and the $v=1, J=5-4$ SiO maser line toward two thermal-pulsating (TP-) AGB stars, R Crt and R Leo. 
The observed fractional linear polarization in the CO emission is measured as $m_l\sim 3.1\%$ and $m_l\sim9.7\%$ for R Crt and R Leo respectively.
A circumstellar envelope (CSE) model profile and the associated parameters are estimated and used as input to a more detailed modeling of the predicted linear polarization \boldred{expected} from the G-K effect. 
%We also utilized a zeroth-order estimate in both circumstellar envelope (CSE) temperature and density profile to give an estimated linear polarization from G-K modeling \citep{yl10}. 
The observed thermal line polarization level is consistent with the predicted results from the G-K model for R Crt; additional effects need to be considered for R Leo.
\section{Introduction}
Key uncertainties remain concerning the morphology and magnitude of magnetic fields in the circumstellar environments (CSEs) around late-type evolved stars \citep{LealF13,lebre14,duthu17}. This includes the relative dynamical influence of magnetic fields in shaping mass-loss outflows from TP-AGB stars relative to competing mechanisms such as wind interaction models and binarity \citep{garcia99,garcia14,matt00,blackman01,soker06,kwok78,frank_etal_1993_ISW,balick_frank_2002rev,soker_04_binary_sum,garcia18,frank18}. 
%{\it Need references here to these questions}

The interstellar magnetic field may be traced by observations of dust continuum polarization \citep{lazarian07,hoanglazarian08}, the Zeeman effect \citep{crutcher_annrev_2012,gkk73,elitzur_III}, and the Goldreich-Kylafis (G-K) effect \citep{gk1,gk2,gk3,dw,lis,cortes}.
In the current work we present continuum and molecular line polarization observations at 1.3mm wavelength with the Combined Array for Research in Millimeter-wave Astronomy (CARMA\footnote{https://www.mmarray.org}) of the CSE toward the TP-AGB stars R Leo and R Crt. Our goal is to use the joint and complementary constraints provided by these three methods to explore the circumstellar magnetic field of late-type evolved stars and to illuminate CSE evolution between the post-AGB and proto-planetary nebulae (PNe) phases.

Dust polarization observations allow the magnetic field morphology to be inferred in the plane of the sky from aligned dust grain emission \citep{lazarian07} but do not provide the field magnitude \citep{crutcher_annrev_2012}. 

The Zeeman effect in both thermal and maser line emission is the primary technique for measuring magnetic field strength directly, particularly for paramagnetic species, as reviewed by \citet{crutcher_kemball_2019}. Given the intrinsic compactness and brightness of maser components, in high angular resolution polarization observations they act as vital probes of the magnetic field at small spatial scales, including shock-enhanced and higher density regions in the CSE \citep{reid_moran_81,gray_book12,moran_etal79,elitzur80,miyoshi_etal94,diamond_etal94}. Molecular species observed include SiO \citep{kemball_diamond97,diamond_kemball_03,kemball09}, H$_{2}$O \citep{fiebig_gusten_89,vlemmings06}, and OH \citep{kemball_diamond_93}. 
Such observations, coupled with a radiative transport model for polarized maser emission, allow the inference of both the projected magnetic field direction and magnitude in principle. In practice there are ambiguities due to uncertainties in polarized maser radiation transport theory \citep{watson09,gray_book12}. 

The G-K effect produces linear polarization in thermal line emission induced by anisotropic velocity gradients that produce an anisotropic radiation field or an intrinsic radiation anisotropy in the presence of a magnetic field $\gtrsim 1 \mu$G \citep{gk1,gk3}. Under the large velocity gradient approximation (LVG) \citep{sobolev60} a velocity gradient produces a gradient in optical depth.  
This yields a net linear polarization in the line emission either parallel or perpendicular to the local magnetic field. 
This ambiguity can be resolved through careful modeling \citep{gk1,cortes,yl10}. 
G-K observational studies have been undertaken toward molecular clouds \citep{lai03,cortes,Li11}, star-forming region outflows and jets \citep{ching16,lee18}, AGB stars \citep{wannier83,glenn97,vlemmings12,girart12}, red supergiant (RSG) stars \citep{vlemmings17}, and PNe \citep{sabin14}. 
The principal advantage of G-K effect observations is probe field morphology in depth; the CSE contains a range of molecular species with varying chemical stratification and excitation profiles over the envelope. The G-K polarization signal also arises from underlying physics that is distinct from the generating mechanisms for continuum dust polarization and the Zeeman effect. It is therefore subject to different observational and theoretical systematic errors.

The measured radial power-law index of the ABG CSE magnetic field strength \citep{reid_etal79,reid_90,LealF13} has been estimated from the Zeeman effect in various molecular species \citep[][and references therein]{herpin06,LealF13,duthu17}. There are competing arguments concerning the magnitude, origin, morphology, and dynamical influence AGB circumstellar magnetic fields \citep{blackman01,thomas95,nordhausblackman06,soker06,soker_zoabi02,soker02d}. \citet{lebre14} report optical spectropolarimetric observations of the S-type Mira star $\chi $ Cyg that infer a surface-averaged magnetic field strength of $\sim 0.5$ G. These authors argue against a global dipole or poloidal morphology due to the implied 
photospheric field strengths ($B \sim 10^{2-3}$ G). In conjunction with optical spectropolarimetry of the non-rotating RSG Betelgeuse \citep{petit13}, they argue for local fields arising from convective dynamos \citep{soker06,dorch04}.

Localized magnetic field enhacement has been proposed above stellar magnetic spots \citep{soker02d} or as a result of pulsation shock compression of the tangential field \citep{hartquist_etal97,kemball09,richter_etal16}. Maser polarization magnetic field strengths imply a magnetic energy density exceeding the thermal and ram pressure \citep{reid_iau_07,richter_etal16}. Further, \citet{watson09} argues that the measured circular SiO maser polarization may arise from non-Zeeman effects. Both factors have been used to argue against the hypothesis of a global dynamically-significant magnetic field. However, recent G-K observations of CO emission toward OH17.7-2.0 \citep{vlemmings2019} suggest a global field morphology consistent with prior OH maser observations. The morphology, magnitude, and dynamical influence of magnetic fields in late-type evolved stars therefore remains a critical open question in both theory and observation \citep{vlemmings2019}.

%%%%%%%%%%%%%%%%%%%% Open questions
The role magnetic fields play in developing axisymmetry from TP-AGB to PNe evolutionary phases is presently unclear \citep{blackman01,nordhausblackman06,garcia99,garcia14}. Outflow collimation \citep{vanmarle14} and direct shaping of the CSE material around isolated AGB stars \citep{Rudinger_hollerbach05,nordhaus07} have both been attributed to magnetic fields. High angular resolution studies of total intensity and magnetic field morphology in AGB stars are needed to further inform this issue.

Target source properties follow and are summarized in Table \ref{tab:table1}, including the right ascension and declination (J2000), chemical classification, variability type, pulsation period $P$ (days), and distance (pc) for each star.

The results presented here are believed to be the only spectral-line CARMA polarimetry of late-type evolved stars at $\lambda =1.3$mm. \boldred{We detect linear polarization in the CO thermal emission at the level $m_l \sim 3-9\%$ which we believe is due to the G-K effect. Associated modeling and analysis are presented.} 

The paper is structured as follows. In Section 2, we describe the observations and data reduction process. The observational results are presented in Section 3. Subsequent modeling, analysis, and discussion are outlined in Section 4 and conclusions are presented in Section 5.

%RCrt
\subsection{R Crt}
\label{sec:rcrt}
R Crt is a semiregular (SRb) AGB star at a \boldred{parallax distance 236 pc \citep{GaiaDR218b,Gaiaparallax18}} with a pulsation period of 160 days \citep{gcvs5} and a mass-loss rate ${\dot M} \sim 8 \times 10^{-7} \left[M_{\odot}/{\rm yr} \right] $ \citep{paladini2017}. \boldred{It has been found that the Gaia parallax results for the extended, pulsating, and bright AGB stars are often unreliable \citep{vanLangevelde_2018,xu19,ramstedt_deathstar_2020}. As discussed in \citet{ramstedt_deathstar_2020} this is even true for stars with what appear to be good astrometric solutions, such as R Crt. However, since R Crt is an SRb variable star, this could also imply significant uncertainties when using a Period-Luminosity (P-L) distance \citep{feast89}; accordingly we adopt the Gaia distance for R Crt. Adopted distances in Section \ref{sec:rcrt} and \ref{sec:rleo} are used only in establishing Figure length scales in this paper.}
The inferred magnetic field strength from single-dish SiO maser Zeeman observations ranges from  $B \sim 0.0-3.7$ G over spectral components \citep{herpin06}.
R Crt has been identified with a companion star \citep{proust1981,cox12}, and is believed to harbor an outflow structure \citep{ishitsuka01,kim_KVN18} possibly aligned with the magnetic field direction inferred from OH maser observations \citep{szymczak99}. 
%

%RLeo
\subsection{R Leo}
\label{sec:rleo}
R Leo is a nearby, typical M-type AGB star \boldred{with a P-L distance of 95 pc \citep{matthews18} derived using \citet{feast89} and data from \citet{haniff_1995}.} It has been intensively studied in several molecular lines including: SiO lines \citep[e.g.][]{cotton09,devicente16,herpin06}, H$_{2}$O masers \citep[e.g.][]{menten91,yates95}, OH masers \citep[e.g.][]{fish06,etoka97}, CO thermal lines \citep[e.g.][]{ramstedt14,debeck10,teyssier06}, and HCN thermal lines \citep{schoier13}. 
The dust shell properties have been observed in infrared and radio continuum \citep[e.g.][]{wittkowski16,ireland04,schoier13}; 
R Leo shows statistically-significant deviations from sphericity in its radio photosphere \citep{matthews18,reid07} and its dusty environment \citep{paladini2017}. 
The pulsation period of R Leo is 310 days \citep{leo_P,gcvs5} and the mass-loss rate estimate is $\dot{M} \sim 1.1 \times 10^{-7} \left[M_{\odot}/{\rm yr} \right] $ \citep{success15}. 
The magnetic field strength in the near CSE of R Leo is estimated from single-dish SiO maser Zeeman observations to be: $B \sim 4.2-4.6$G \citep{herpin06}. R Leo has also been suggested as hosting a Jovian planet remnant \citep{wiesemeyer09}. 
Recent ALMA observations of R Leo \citep{vlemmings_etal19} provide important insight into the extended atmosphere region $R\sim 1-2R_*$ and measure an asymmetric expansion with a mean velocity of $10.6\pm1.4\ {\rm km/s}$.

\begin{table}[ht!]
  \centering
  \caption{Source information}
  \label{tab:table1}
  \begin{tabular}{c||cccccc}
  \hline
    Name & RA & Dec & Chem. class & Variability  & P & Distance \\
    {} & (J2000) & (J2000) & {} & {} & (days) & (pc)\\
%    (reference) & -- & -- & (a) & (a) & (a) & (b)\\
    \hline
    R Crt & 11:00:33.85 & -18:19:29.85 & O-rich$^{(a)}$& SRb$^{(a)}$ & 160$^{(a)}$ & \boldred{$236^{(b)}$} \\
    \hline
    R Leo & 09:47:33.49 & +11:25:43.67 & O-rich$^{(a)}$& Mira$^{(a)}$  & 309.95$^{(a)}$ & \boldred{$95^{(c)}$} \\
    \hline
  \end{tabular}
    \begin{tablenotes}
  	\small
  	\centering
    \item{a. Chemical classification, variability type, and pulsation period $P$ are from \citet{gcvs5}.}
   \boldred{\item{b. Parallax distance estimate is from Gaia DR2 \citep{GaiaDR218b,Gaiaparallax18}.}}
    \item{c. P-L distance estimate is from \citet{matthews18}.}
  \end{tablenotes}
\end{table}

\section{Observation and data reduction}
\subsection{Observation}
The observations of R Crt and R Leo were performed in the  CARMA E-configuration ($8.5 - 66$ m baseline lengths). Two observing runs were performed, on November 16, 2014 and November 11, 2014 respectively, with each run of total duration 3.7 hours.
The correlator was configured in full Stokes mode to observe the SiO $v=1, J=5-4$ line in the lower sideband (LSB) and the CO $J = 2-1$  line in the upper sideband (USB), with adopted transition rest frequencies of 215.596 GHz and 230.538 GHz respectively \citep{lovas_2009}. Both lines were observed in narrow-band (line) spectral windows of bandwidth $\triangle \nu = 62.2$ MHz each sampled over $191$ frequency channels with corresponding nominal velocity resolutions of $0.453$ km/s and $0.423$ km/s respectively. In addition, two wide-band (continuum) spectral windows each of bandwidth $\triangle \nu = 489.6$ MHz and sampled over 47 frequency channels each were centered on the spectral lines.

%The table of pertinent information about the sources is listed below, with a brief summary of the details of the sources from the \textit{General Catalogue of Variable Stars}, GCVS5, \citep{gcvs5}  in Table \ref{tab:table1}. 
\subsection{Data reduction}
We describe the data reduction using the general interferometric data model developed by \citet{hamaker96_I} and \citet{sault96_II} adopting a circularly-polarized basis as used by CARMA. Here, $2\times 2$ Jones calibration matrices $\mathbf{J}_{m} \in \{\mathbf{G}_{m}, \mathbf{B}_{m}, \mathbf{D}_{m},...\}$ are used to represent all linear instrumental and propagation signal-path effects at each antenna $m$ connecting the incident electric field $\mathbf{E}_{m}=[E^R_m, E^L_m]^T$ to the measured signal $\mathbf{E}^{\prime}_{m}$. Jones matrix $\mathbf{G}_{m}$ defines the net frequency-independent gain, $\mathbf{B}_{m}$ the net frequency-dependent passband, and $\mathbf{D}_{m}$ the instrumental polarization leakage. The radio-interferometric cross correlation on baseline $m$-$n$ takes the form:
\begin{align}
    \mathbf{R}^{\prime}_{mn} = \langle \mathbf{E}^{\prime}_{m} \otimes \mathbf{E}^{\prime *}_{n} \rangle = \left(\prod \mathbf{J}_{m} \otimes \mathbf{J}^{*}_{n}\right) \mathbf{R}_{mn} \nonumber \\
    &= \left[ \left( \mathbf{G}_{m}\mathbf{B}_{m}\mathbf{D}_{m} \cdots \right) \otimes \left( \mathbf{G}^{*}_{n}\mathbf{B}^{*}_{n}\mathbf{D}^{*}_{n} \cdots \right) \right] \mathbf{R}_{mn} 
    %\\&= \left[ \left( \mathbf{G}_{m}\otimes \mathbf{G}^{*}_{n} \right) \left( \mathbf{B}_{m}\otimes \mathbf{B}^{*}_{n} \right) \left( \mathbf{D}_{m}\otimes \mathbf{D}^{*}_{n} \right) \cdots \right]\mathbf{R}_{mn}
\end{align}
where $\otimes$ denotes the outer matrix product and $\mathbf{R}_{mn}$ is the intrinsic coherence. 
CARMA continuum polarimetry is described by \citet{hullplambeck}. Polarization calibration for the pre-cursor Owens Valley Radio Observatory (OVRO) is described by \citet{akeson97_phdt}. 

The parallel-hand (RR,LL) data calibration was performed using the CArma Data REduction pipeline, $\mathtt{CADRE}$ \citep{cadre}, based on the Multichannel Image Reconstruction, Image Analysis and Display analysis package ($\mathtt{MIRIAD}$) \citep{miriad}. 
The parallel-hand reduction corrects the passband ($\mathbf{B}_{m}$) and gain ($\mathbf{G}_{m}$) calibration terms. The $\mathtt{CADRE}$ pipeline was extended to support full-polarization reduction. For R Crt, 3C279 was used as the bandpass, gain, and leakage calibrator. For R Leo, 3C84 was used as the bandpass calibrator, and OJ287 as the gain and leakage calibrator. The gain calibrator absolute flux densities were obtained from high-quality ALMA band 6 ($211 - 275$ GHz) calibrator catalogue\footnote{https://almascience.nrao.edu/alma-data/calibrator-catalogue} entries adjacent in time. The resulting flux densities were $ 8.38 \pm 0.39 $ Jy (3C279) and $ 3.04 \pm 0.17 $ Jy (OJ287). 
The antenna-based R and L gains $\mathbf{G}_{m}={\rm diag}(G^{R}_{m}, G^{L}_{m})={\rm diag}(g^R_m e^{i\phi_m^R}, g^L_m e^{i\phi_m^L})$ were solved separately over time using the RR and LL data respectively. The underlying data were flagged for gain solutions showing high time-variability or outlier behavior in the differential polarization amplitude gain ratio $g^{R/L}_{m}$. 

Full polarization calibration requires augmentation of parallel-hand reduction by: i) R-L phase ($\phi_{RL,m}= \phi_{m}^{R} - \phi_{m}^{L}$) calibration; and ii) polarization leakage (D-term, $\mathbf{D}_{m}$) calibration \citep{TMS_2017,hullplambeck}. 
At millimeter wavelengths $\phi_{RL,m}$ is primarily instrumental as the atmosphere is predominantly non-dispersive \citep{kemball_richter_2011}.
The CARMA 10-m telescopes allow injected calibration of the R-L phase \citep{Hull_14PhDT,hullplambeck}. 
The $\mathtt{MIRIAD}$ task $\mathtt{xyauto}$ solves for the R-L phase corrections for all the 10-m antennas using this method
\citep{Hull_14PhDT,hullplambeck}. 
The R-L solution from a reference 10-m antenna was then bootstrapped and applied to the 6-m antennas to correct their R-L phases using $\mathtt{MIRIAD}$ task $\mathtt{mfcal}$. 
The instrumental polarization leakage term $\mathbf{D}_{m}$ is solved for from the cross-polarized data (RL and LR) obtained on a bright calibrator source after parallel-hand calibration and correction for the residual R-L phase difference, for observations over a sufficient range $\Delta \alpha$ of parallactic angle coverage \citep{TMS_2017}. 
The $\mathbf{D}_{m}$ terms were solved for using the $\mathtt{MIRIAD}$ task $\mathtt{gpcal}$ with $\mathtt{options=qusolve}$. 

\begin{figure}
  \centering
  \begin{tabular}[b]{@{}p{0.45\textwidth}@{}}
    \centering\includegraphics[width=1.0\linewidth]{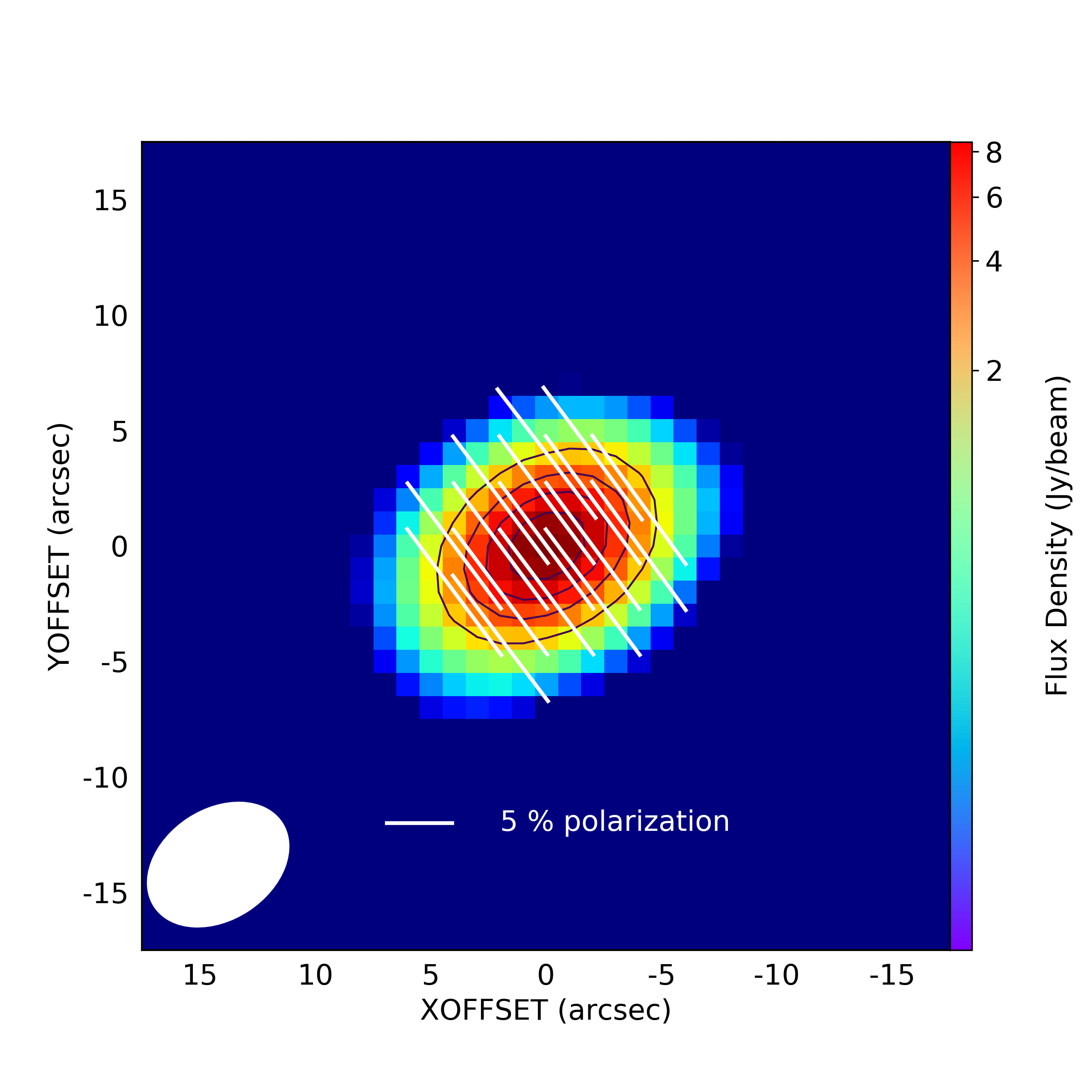} \\
    \centering\small (a) Polarization map of 3C279, LSB
  \end{tabular}%
  \quad
  \begin{tabular}[b]{@{}p{0.45\textwidth}@{}}
    \centering\includegraphics[width=1.0\linewidth]{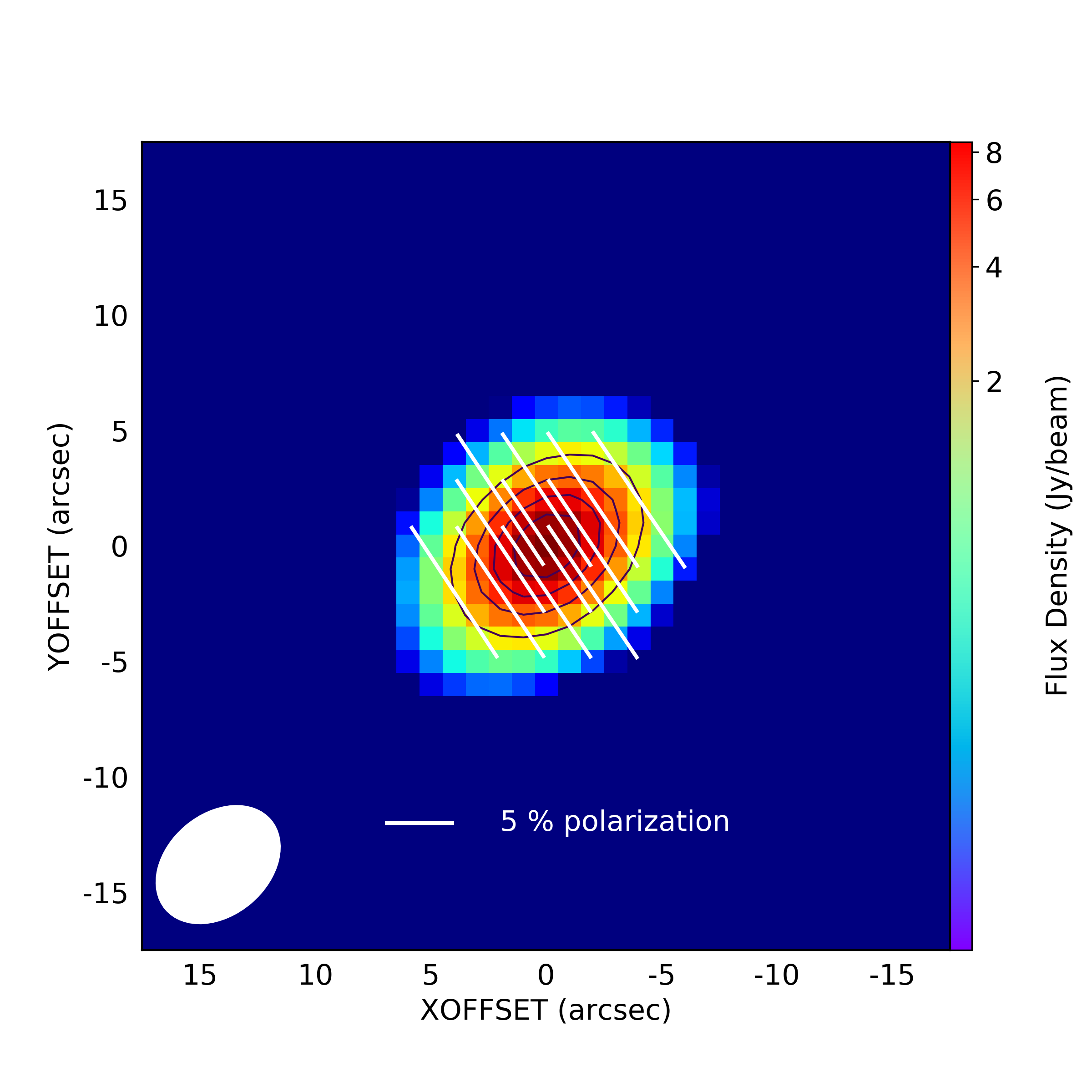} \\
    \centering\small (b) Polarization map of 3C279, USB
  \end{tabular}
  \caption{Linear polarization maps of 3C279 in both LSB and USB. The color map shows the total intensity; the peak total intensity is $8.361$ Jy/beam in LSB, and $8.363$ Jy/beam in USB. The contour levels as the linear polarization intensity from 20\%, 40\%, 60\%, and 80\% of the maximum linear polarization intensity, which is $0.968 $ Jy/beam in LSB, and $0.967 $ Jy/beam in USB respectively. The synthesized beam is shown at lower left.} %\AKnote{need to cite the maximum polarized intensity; also note the color wedge at right for the Stokes I brightness - does miriad label this flux density instead of brightness ?} }
  \label{fig:3c279map}
\end{figure}

\begin{figure}
  \centering
  \begin{tabular}[b]{@{}p{0.45\textwidth}@{}}
    \centering\includegraphics[width=1.0\linewidth]{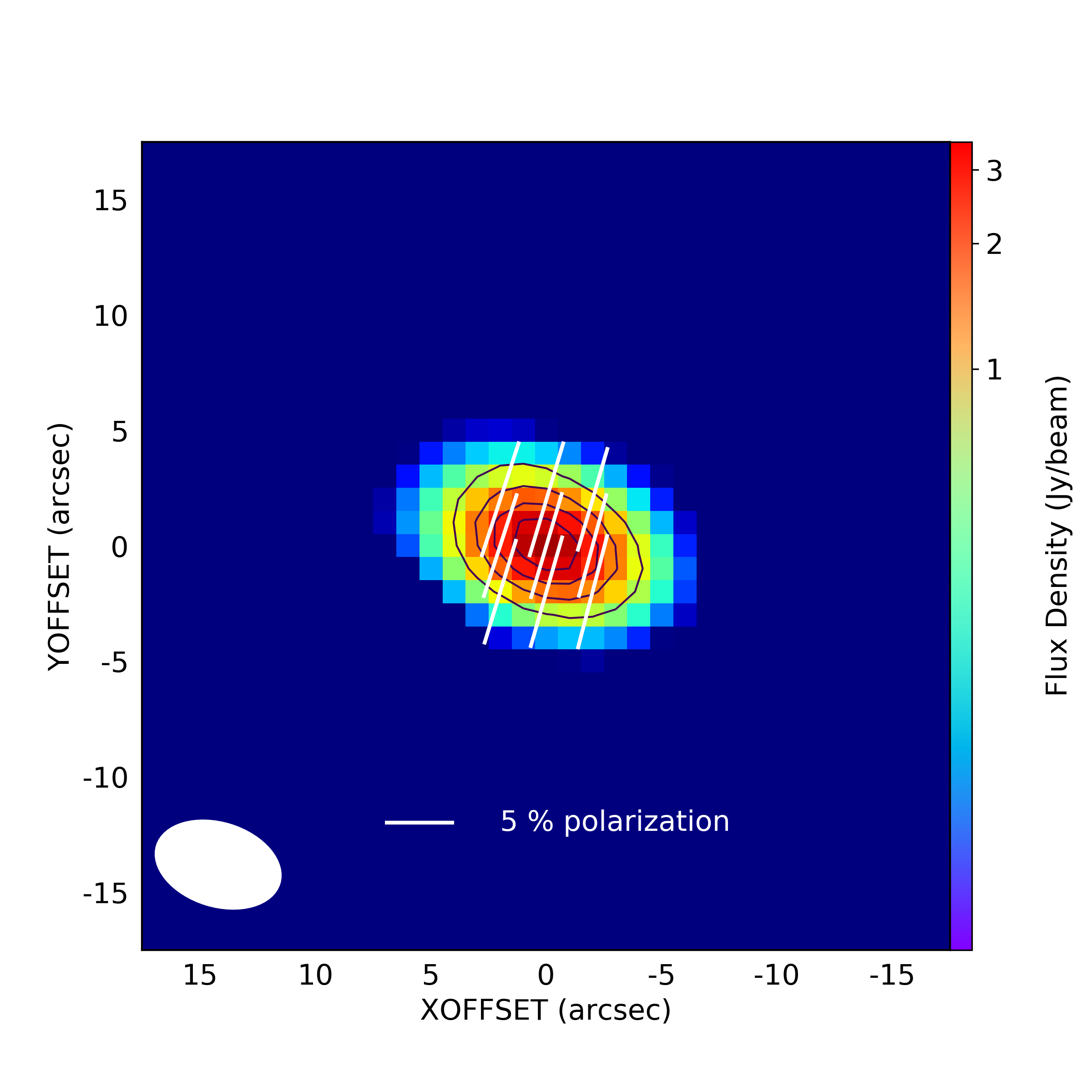} \\
    \centering\small (a) Polarization map of OJ287, LSB
  \end{tabular}%
  \quad
  \begin{tabular}[b]{@{}p{0.45\textwidth}@{}}
    \centering\includegraphics[width=1.0\linewidth]{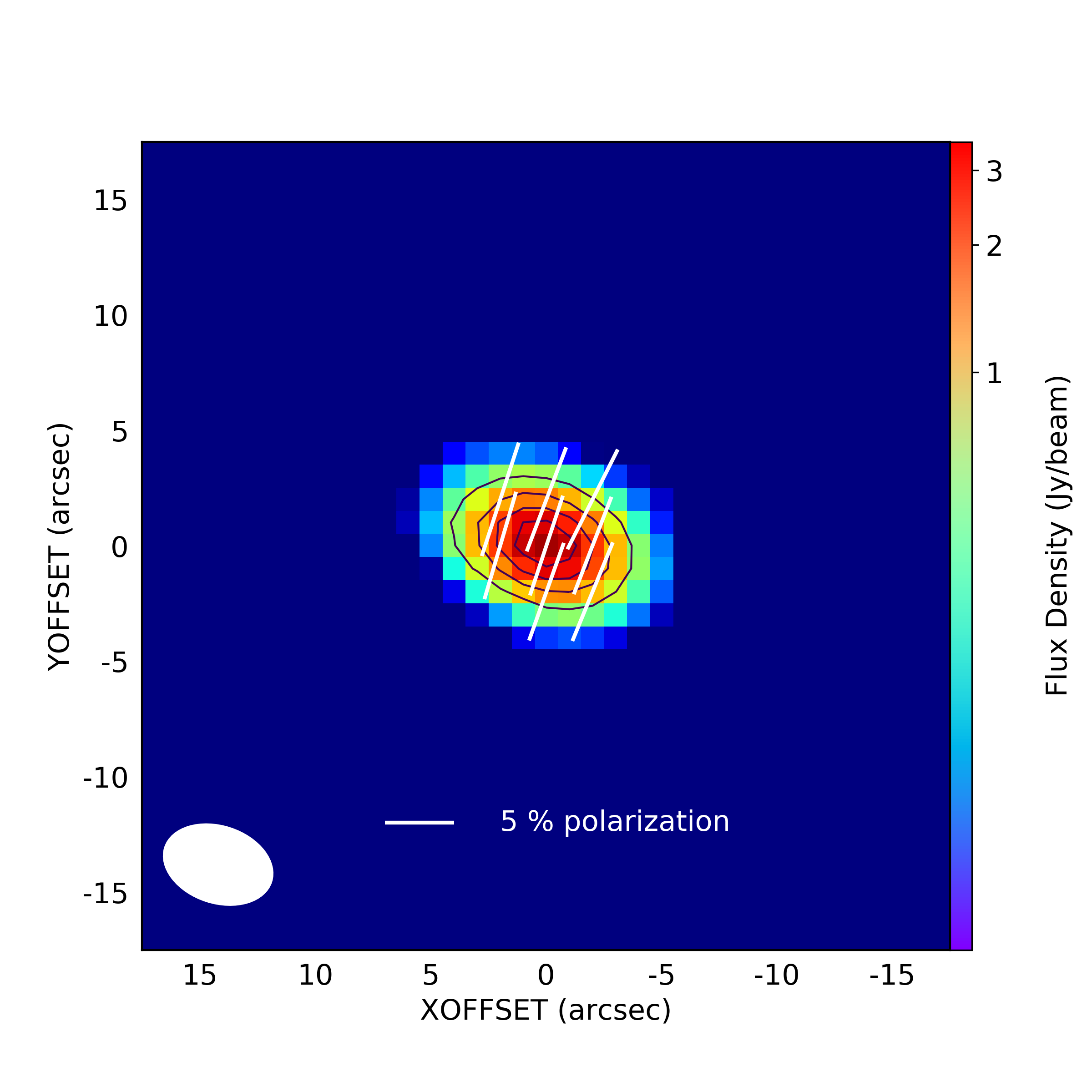} \\
    \centering\small (b) Polarization map of OJ287, USB
  \end{tabular}
  \caption{Linear polarization maps of OJ287 in both LSB and USB. The color map shows the total intensity; the peak total intensity is $3.034$ Jy/beam in LSB, and $3.029$ Jy/beam in USB. The contour levels as the linear polarization intensity from 20\%, 40\%, 60\%, and 80\% of the maximum linear polarization intensity, which is $0.244 $ Jy/beam in LSB, and $0.230 $ Jy/beam in USB respectively.The synthesized beam is shown at lower left.}%\AKnote{same comments as Figure 1} }
  \label{fig:oj287map}
\end{figure}

The IQUV images were formed from the fully calibrated visibilities by $\mathtt{MIRIAD}$ task $\mathtt{invert}$ using natural visibility gridding weighting.
At $230$ GHz, the beam size is $5.73'' \times 4.21''$ (at PA $=29.7^{\circ}$) for the calibrated R Crt data and $5.07'' \times 3.58'' $ (at PA = $80.1^{\circ}$) for the calibrated R Leo data. 
The synthesized maps of linearly polarized intensity $P$ were de-biased \citep{wardle_kronberg_74} using $\mathtt{MIRIAD}$ task $\mathtt{impol}$ and were masked below $5 \sigma$ in Stokes $I$ and $4 \sigma$ in $P$. 
The spectral channels were averaged over 1 km s$^{-1}$ in both the CO and SiO channel maps. The final calibrated continuum maps of the gain calibrators, 3C279 and OJ287, are shown in Fig.\ref{fig:3c279map} and Fig.\ref{fig:oj287map}. 
The calculated theoretical noise level in the spectral-line image cubes for a 1.0 km s$^{-1}$ channel width is between $29.6$ mJy/beam (good weather) $\sim 56.8$ mJy/beam (typical weather)\footnote{http://bima.astro.umd.edu/carma/observing/tools/rms.html}.
The measured median of the off-source noise over frequency from the spectral-line, total intensity data cubes is between $50.1 \sim 274.1$ mJy/beam, over the separate transitions. 

Several polarization and calibration quality assurance tests were applied to the fully calibrated data; these are described in Appendix \ref{sec:qa}.


\begin{table}
  \centering
  \caption{Summary of the detected molecular lines of R Crt and R Leo in CARMA observation. }
  \label{tab:table2}
  \begin{tabular}{ccccc}
  \hline
  \hline
   Source & Line & $\nu_{0}$ & F$_{\ell}$ & F\textsubscript{I}  \\
    {} & {} & [GHz] &[ Jy beam$^{-1}$  & [Jy beam$^{-1}$ \\
    {} & {} & {} & km s$^{-1}$]& km s$^{-1}$]\\
    \hline
%    RCrt & {} & {} & {} & {}
     R Crt & CO (J = 2-1)        & 230.538  & 3.60 & 433.9 \\
     {} & SiO (J=5-4, v=1)       & 215.596  & 1.31 & 27.9 \\
    \hline
%    RLeo & {} & {} & {} & {}
     R Leo & CO (J = 2-1)        & 230.538   & 2.62 & 297.9  \\
     {} & SiO (J=5-4, v=1)       & 215.596   & 89.9 & 492.3 \\
    \hline
  \end{tabular}
  \begin{tablenotes}
  	\small
    \centering
    \item Note: The listed are the rest frequency ($\nu_{0}$) of the detected molecular lines, the integrated brightness over velocity channels of the linearly polarized intensity (F$_{\ell}$) and total intensity (F\textsubscript{I}) in each observation
  \end{tablenotes}
  %\AKnoteM{not sure why I listed F $[$Jy/beam$]$ $[$km/s$]$}
  %\AKnote{Need to define the columns in a Table footnote. Believe rest frequency used for SiO was 215.596 GHz}
\end{table}
\section{Results}
\begin{figure}
  \centering
  \begin{tabular}[b]{@{}p{0.45\textwidth}@{}}
    \centering\includegraphics[width=1.0\linewidth]{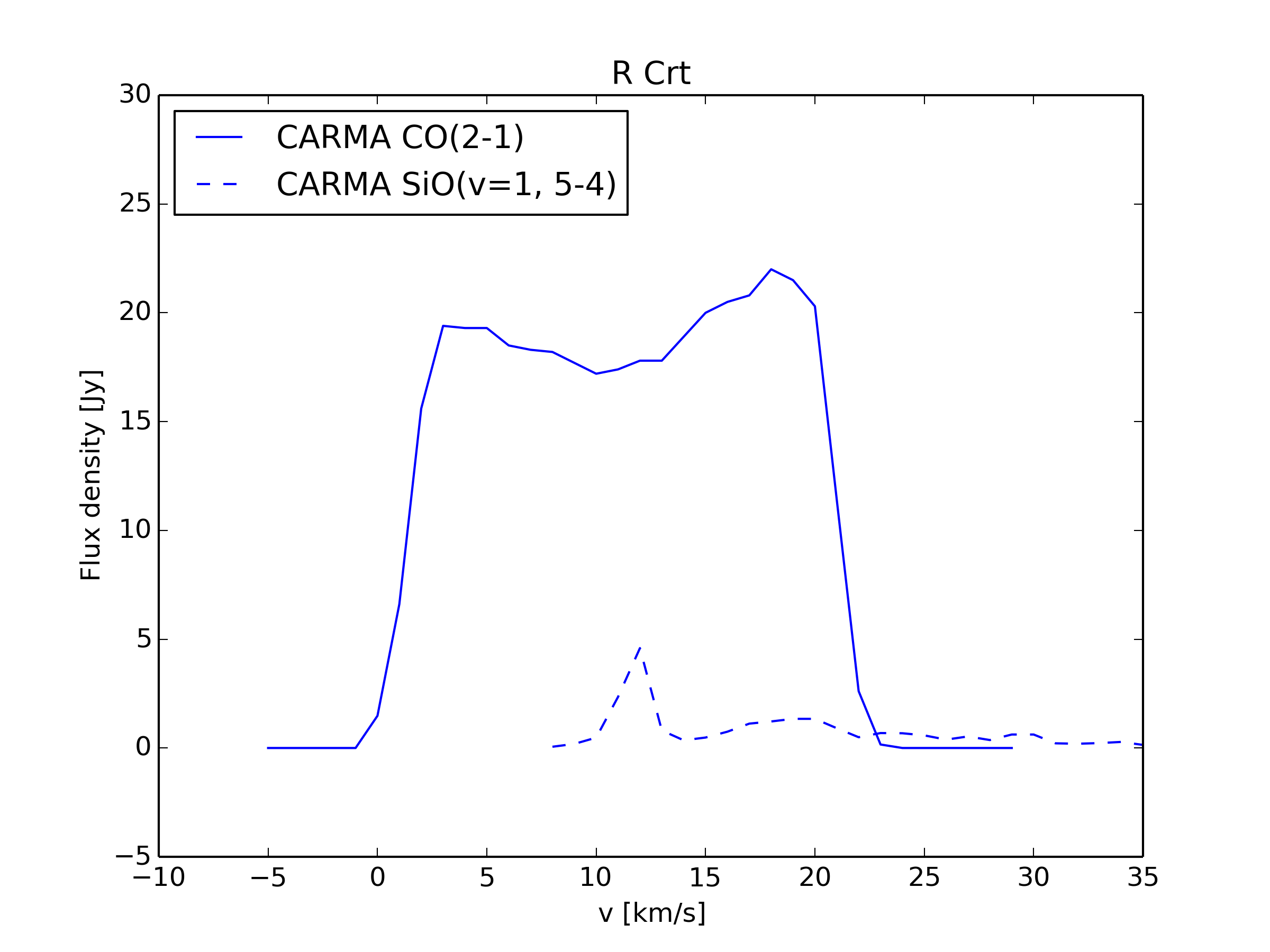} \\
    \centering\small (a) R Crt
  \end{tabular}%
  \quad
  \begin{tabular}[b]{@{}p{0.45\textwidth}@{}}
    \centering\includegraphics[width=1.0\linewidth]{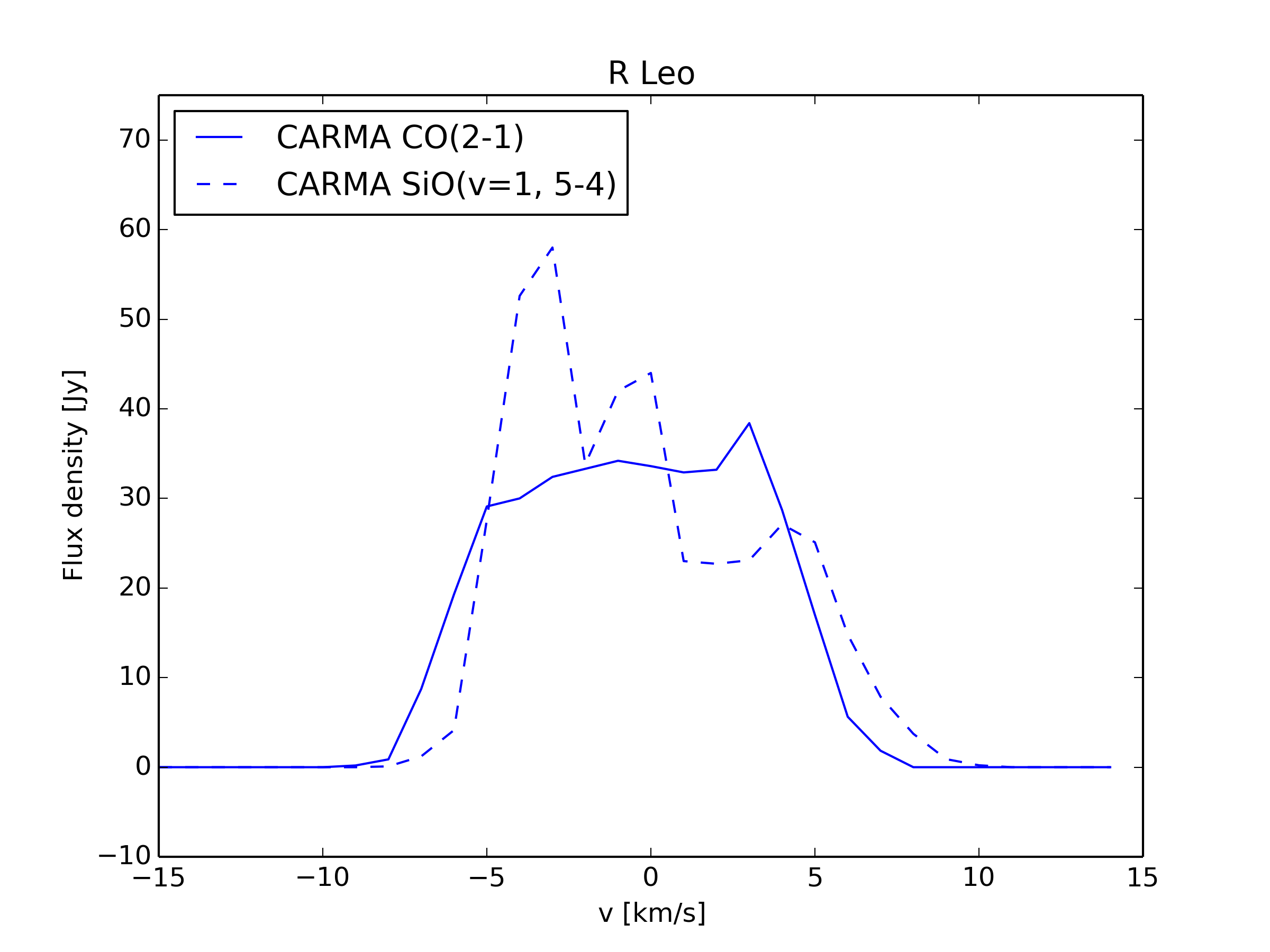} \\
    \centering\small (b) R Leo
  \end{tabular}
  \caption{Integrated spectra for both (a) R Crt, and (b) R Leo. The spectra were obtained by summing the total intensity at all points above $3\sigma$ in each image plane of the interferometric image cube using $\mathtt{MIRIAD}$ task $\mathtt{imspec}$. The spectra are binned at a $\triangle v= 1 {\rm km/s}$ interval.}
%  \caption{Emission spectra of both (a)R Crt and (b)R Leo. These are the averages of cross-power spectra across all available baselines. }
  \label{fig:spec}
\end{figure}

In Table \ref{tab:table2} the rest frequency $\nu_{0}$ and the linearly-polarized intensity F$_{\ell}$ and total intensity F\textsubscript{I} integrated over velocity are enumerated per source and molecular line. Total power spectra obtained by summing all total intensity points above $3\sigma$ in each image plane of the interferometric image cubes are shown in Fig.\ref{fig:spec}. 
 
We detect continuum emission toward R Crt of $14.9 \pm 2.4$ mJy/beam in the LSB continuum spectral window and $18.3\pm 2.9$ mJy/beam in the counterpart USB window. These values are measured at the peak-brightness pixel and the error is the off-source rms noise.  In the same spectral windows, continuum emission is detected toward R Leo of $118.2\pm 3.2$ mJy/beam in LSB and $123.0\pm 3.4$ mJy/beam in USB using the same measurement methodology. No continuum polarization was detected toward either R Crt or R Leo, likely due to sensitivity limitations and the expected weakness of the dust continuum emission relative to the underlying stellar continuum. \boldred{By comparison with the stellar continuum flux densities reported by \citet{vlemmings_etal19} for R Leo, the dust contribution does not exceed $\sim 20\%$; a comparable result is expected for R Crt.}
   
\begin{figure}
  \centering
  \begin{tabular}[b]{@{}p{0.45\textwidth}@{}}
    \centering\includegraphics[width=1.0\linewidth]{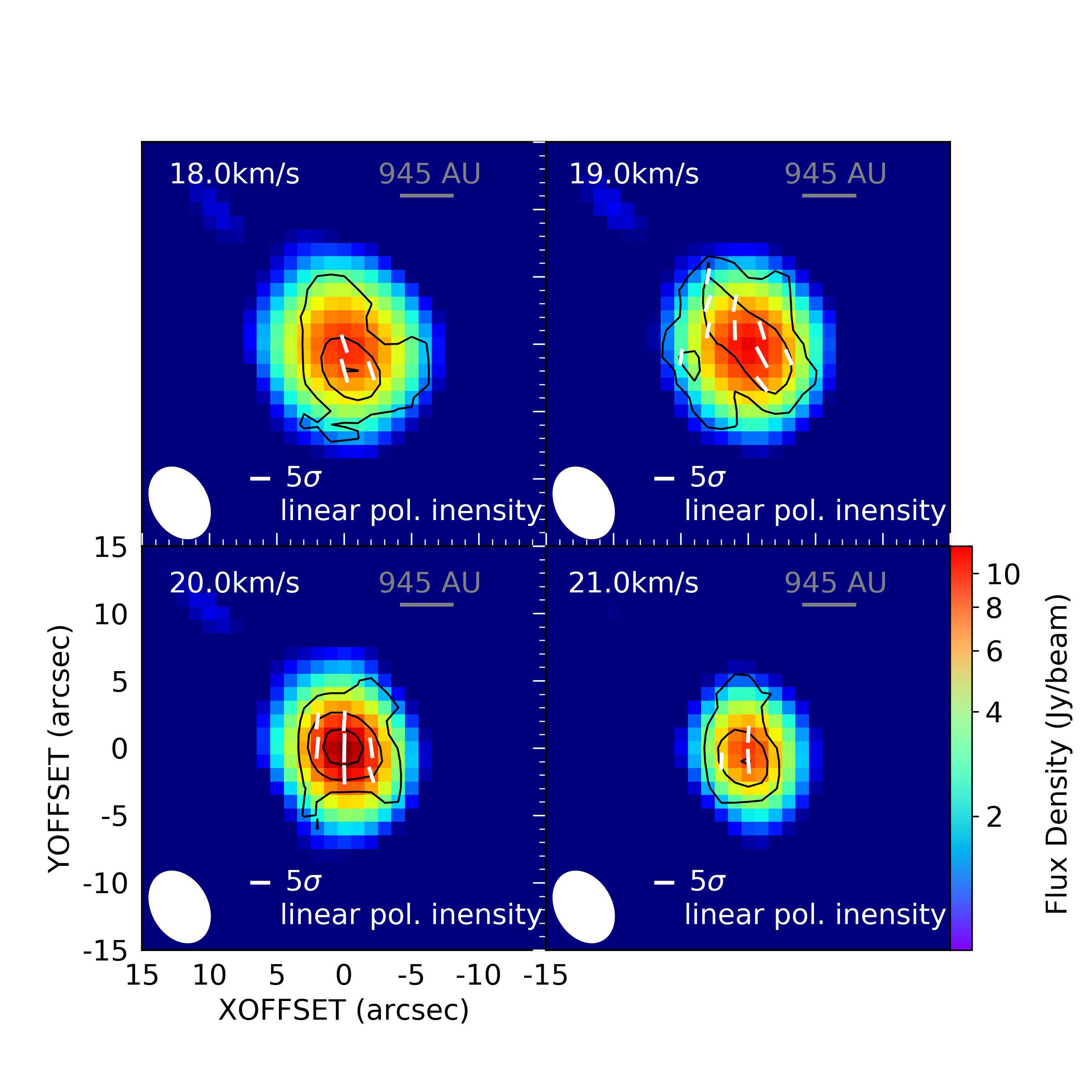} \\
    \centering\small (a) R Crt
  \end{tabular}%
  \quad
  \begin{tabular}[b]{@{}p{0.45\textwidth}@{}}
    \centering\includegraphics[width=1.0\linewidth]{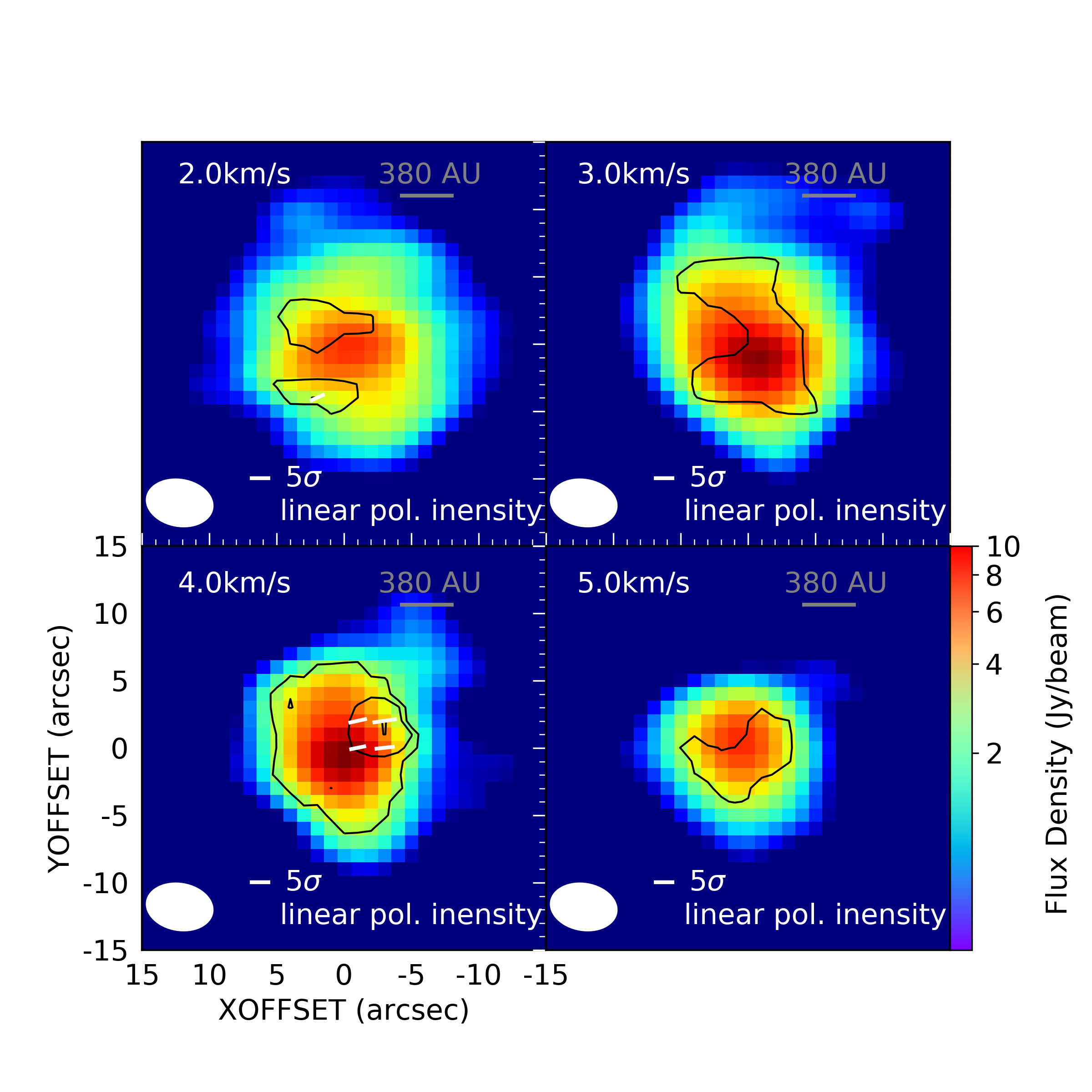} \\
    \centering\small (b) R Leo
  \end{tabular}
  \caption{CO(2-1) spectral-line polarization maps of  (a) R Crt, and (b) R Leo. The channels are averaged over interval of 1 km s\textsuperscript{-1}. The color map shows the spectral-line emission total intensity in log scale, and the contour levels of the linear polarization intensity are \boldred{2, 4, and 6} times $\sigma$ where $\sigma$ is the noise level of each source. The white segments \boldred{are drawn at the EVPA orientation with scaled length proportional to linearly-polarized intensity}. Beam size is also displayed at the lower-left corner in each map.\boldred{The physical scale is derived from the adopted distances in Table \ref{tab:table1}.}}
  \label{fig:carmaco}
\end{figure}

\begin{figure}
  \centering
  \begin{tabular}[b]{@{}p{0.45\textwidth}@{}}
    \centering\includegraphics[width=1.0\linewidth]{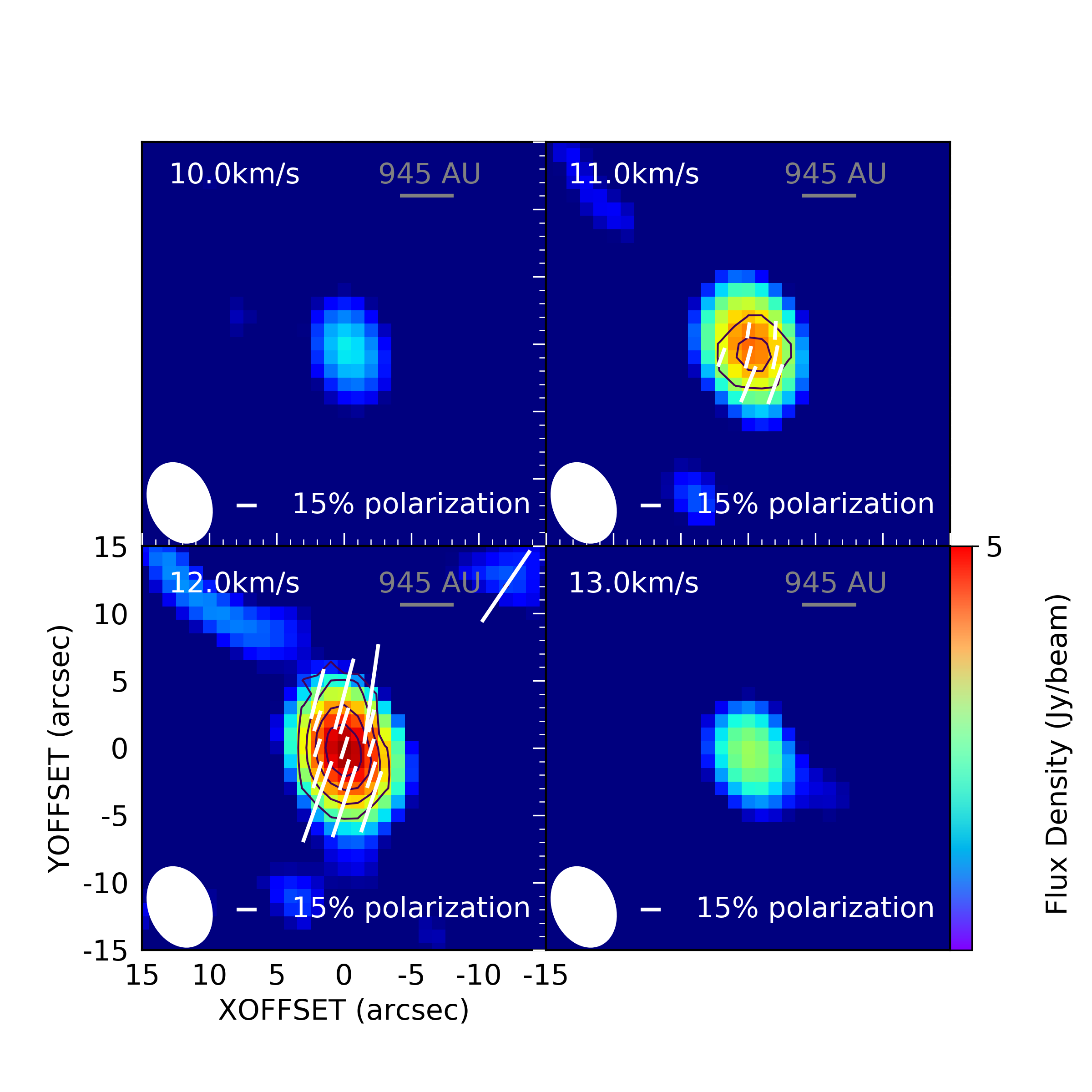} \\
    \centering\small (a) R Crt
  \end{tabular}%
  \quad
  \begin{tabular}[b]{@{}p{0.45\textwidth}@{}}
    \centering\includegraphics[width=1.0\linewidth]{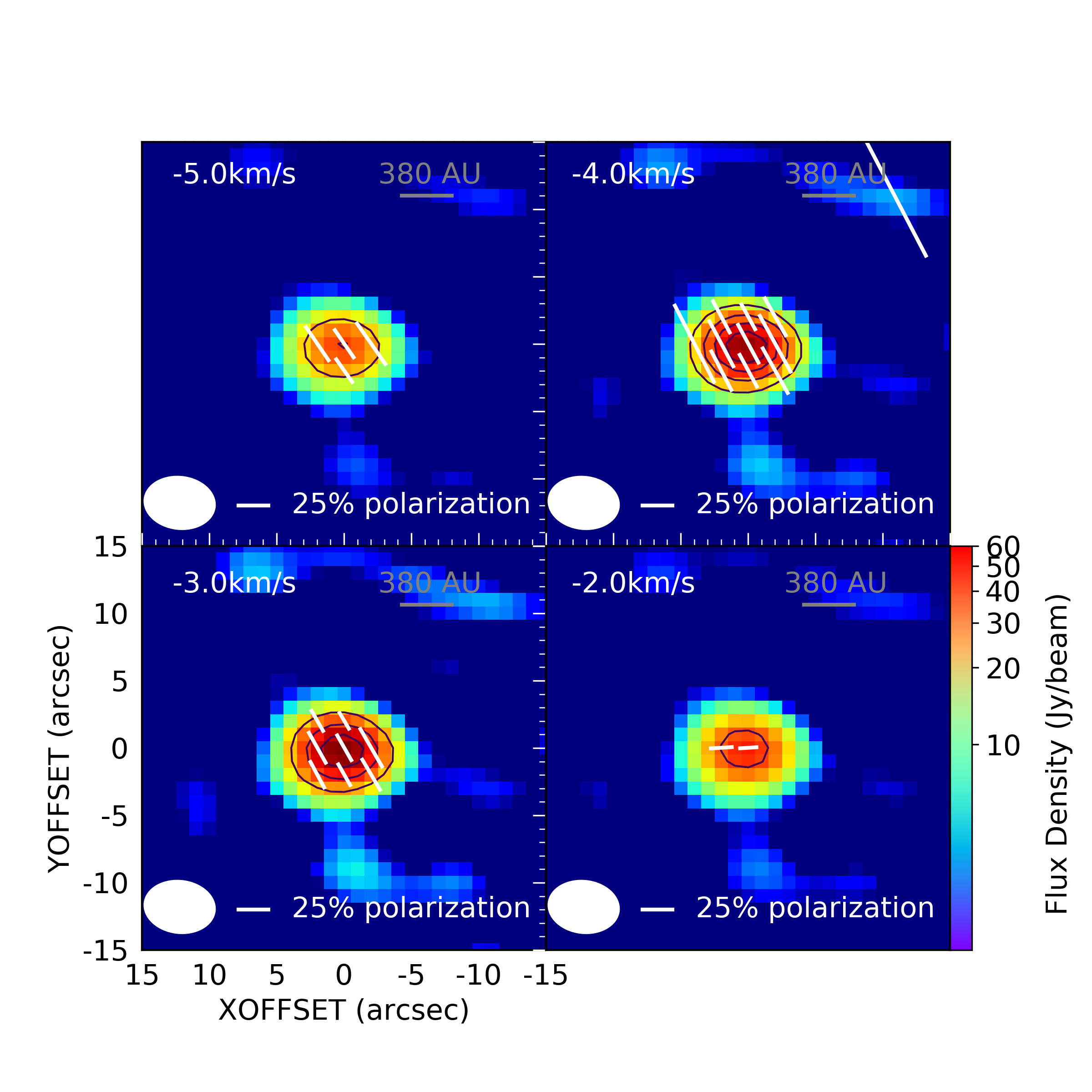} \\
    \centering\small (b) R Leo
  \end{tabular}
  \caption{SiO(5-4, v=1) spectral-line polarization maps of  (a) R Crt, and (b) R Leo. Again the channels are averaged over 1 km s\textsuperscript{-1} interval. The color corresponds to total intensity of the emission in log scale, contour levels as the linear polarization intensity from $20\%$, $40\%$, $60\%$, and $80\%$ of the maximum, and white segments as the polarization angle with the scaled length as display of the linear polarization strength.\boldred{The physical scale is derived from the adopted distances in Table \ref{tab:table1}.} }
  \label{fig:carmasio}
\end{figure}

The $J=2-1$ CO emission channel maps for both sources are displayed in Fig. \ref{fig:carmaco}. Contours depict linearly-polarized intensity $P$ with associated vectors indicating EVPA $\chi$; both are overlaid on color plots of total intensity. Analogous channel maps for $v=1, J=5-4$ SiO emission are shown in Fig. \ref{fig:carmasio}. The peak-pixel linearly-polarized intensity $P$ across these spectral-line image cubes is tabulated in Table \ref{tab:table3} for each source and transition. We denote this as $P_p$. This table includes the associated LSR velocity $v_{LSR}$, point total intensity $I_p$ and resulting fractional linear polarization $m_{l,p}$ at $P_p$. The error-weighted EVPA average ${\bar \chi}$ is computed over the angular ($\sim 10$ arcsec) and frequency extent ($\triangle v\sim 2-4$ km/s) of the source for all pixel detections above $4\sigma$. The statistical error $\sigma_n$ in ${\bar \chi}$ was estimated by propagating the pixel error $\sigma_{\chi}=\sigma_P/{2P}$ \citep{wardle_kronberg_74} over the source summation \citep{error_analysis}. A conservative estimate of the systematic error ($\sigma_{s}$) in ${\bar \chi}$ was obtained from the absolute EVPA comparison against external measurements described in Appendix \ref{sec:qa}. Based on the alignment uncertainty for 3C279, $\sigma_s$ for R Crt is estimated as $10.8^{\circ}$; similarly, via OJ287, $\sigma_s$ for R Leo is estimated as $11.6^{\circ}$. \boldred{The row $\sigma$ in Table \ref{tab:table3} is column-specific, namely $\sigma_P$ for $P_p$, $\sigma_I$ for $I_p$, and takes the format $\pm \sigma_n \pm \sigma_s$ for ${\bar \chi}$.}

\boldred{Table \ref{tab:table3a} lists the peak-pixel circularly-polarized intensity $V_p$ in the $v=1, J=5-4$ SiO spectral-line image cube (Fig. \ref{fig:carmasio}) and associated LSR velocity $v_{LSR}$, point total intensity $I_p$ and resulting degree of fractional circular polarization $m_{c,p}$ at $V_p$; circular polarization is only detected toward R Leo.}

\begin{table}[ht!]
  \centering
  \caption{Polarization properties at the position of peak linearly-polarized intensity in the spectral-line image cubes.}
  \label{tab:table3}
  \begin{tabular}{ccccccc}
  \hline
  \hline
    Source & Line & \boldred{v$_{LSR}$} & $P_p$ & $I_p$ & m$_{\ell,p}$ & 
    ${\bar \chi}$ \\
    {} & {} & [km s\textsuperscript{-1}] & [ Jy beam\textsuperscript{-1}] & [Jy beam\textsuperscript{-1}] & [$\%$]  & [$^{\circ}$]\\
    \hline
    R Crt    & SiO (J=5-4, v=1)  & 12.0 & 0.778     & 4.52    &  17.2 (17.3 $\sigma$) & $-15.4$ \\
    {}    &       $\sigma$    &  & 0.045  & 0.05  & --                    & $\pm 0.7\pm 10.8$\\
    {}    & CO (J = 2-1)  & 20.0    & 0.361     & 11.44   & 3.13 (7.2 $\sigma$)   & $7.6$ \\
    {}    &       $\sigma$    &  & 0.050   & 0.27  & --                    & $\pm 1.0\pm 10.8$\\
    \hline
    R Leo    & SiO (J=5-4, v=1)  & -4.0 & 18.67     & 53.98   &  34.6 (317 $\sigma$)  & $-1.1$\\
    {}    &       $\sigma$    &  & 0.06   & 0.17  & --                    & $\pm 0.03 \pm 11.6$\\
    {}    & CO (J = 2-1)      & 4.0 & 0.375     & 3.86   & 9.71 (6.1$\sigma$)    & $-77.7$\\
    {}    &       $\sigma$    &  & 0.062   & 0.14  & --                    & $\pm 1.7 \pm 11.6$\\
    \hline
  \end{tabular}
\end{table}

\boldred{
\begin{table}[ht!]
  \centering
  \caption{Polarization properties at the position of peak circularly-polarized intensity in the spectral-line image cubes.}
  \label{tab:table3a}
  \begin{tabular}{cccccc}
  \hline
  \hline
    Source & Line & $v_{LSR}$ & $V_p$ & $I_p$ & $m_{c,p}$\\ 
    {} & {} & [km s\textsuperscript{-1}] & [ Jy beam\textsuperscript{-1}]  & [Jy beam\textsuperscript{-1}] & [$\%$]\\
    \hline
    R Crt    & SiO (J=5-4, v=1)  & -- & -- & -- & --\\
    {}    &      $\sigma$    &   -- & -- & -- & --\\
    \hline
    R Leo    & SiO (J=5-4, v=1)  & -3.0 & 2.95 & 51.34 & 5.7 (40.3 $\sigma$)\\
    {}    &       $\sigma$   &  &  0.07 & 0.17 & --\\
    \hline
  \end{tabular}
\end{table}
} 

\section{Analysis and Discussion}
%%%The observed linear polarization of both sources R Crt and R Leo are displayed 
As shown in Table \ref{tab:table3}, the peak fractional linear polarizations for the $J=2-1$ CO line emission are $m_l=3.13\%$ and $m_l=9.7\%$ for R Crt and R Leo respectively. In this section we describe numerical modeling to determine if this linear polarization can be ascribed to the G-K effect. Our modeling of the G-K effect in circumstellar CO requires as input radial CSE profiles in temperature $T(r)$ and density $n(r)$ in addition to a characterization of radiation anisotropy and velocity gradient structure; our adopted CSE model is described in Appendix \ref{sec:cse}. In this section we also consider 
comparisons between the G-K signal, the SiO maser polarization, and further source properties in the literature.

\subsection{G-K modeling}
\subsubsection{Framework}
\begin{figure}
	\center
	\includegraphics[scale=0.5]{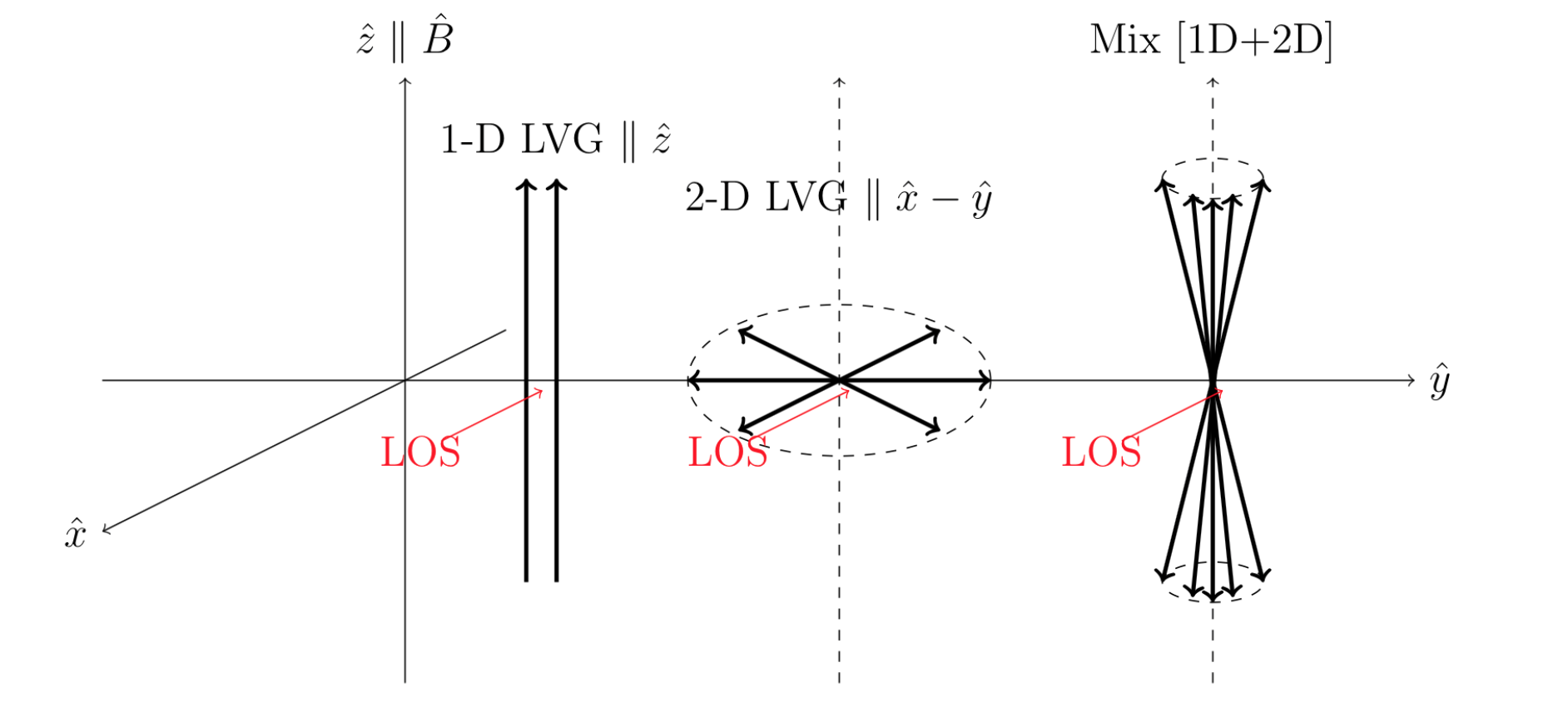}
	\caption{The schematic plot for three types of velocity gradient geometry as implemented in our existing G-K model (from left to right):  i) 1-D velocity anisotropy parallel the magnetic field; ii) a 2-D velocity gradient along the 2-D plane perpendicular to the magnetic field;  and,  iii) a mixed model of (i) and (ii),  yielding a cone-shaped velocity gradient lobe. The "LOS" label in red denotes the line-of-sight direction. The magnetic field orientation $\hat{B}$ is set to be aligned with the z-axis, i.e. $ \hat{z} \parallel \hat{B}$. And the black arrows in the plot point in the direction of the velocity gradient field. In our model we have adopted the mixed LVG velocity gradient geometry on the basis that of the three velocity gradient geometries, this model is closest to the modest axisymmetry that could reasonably be expected in the late AGB phase. }
	\label{fig:vgradient_geo}
\end{figure}

%[assumption: uniform B to the first order, uniform density slab...? and whatever]
We use the G-K modeling framework developed independently by \citet{yl10}. The differences between the numerical code and prior approaches is described briefly here. Original work by \citet{gk1,gk2} and \citet{kylafis83_ano} considered 1-D and 2-D LVG geometries, a $J=1-0$ transition, and a collisional rate $C$ to Einstein coefficient $A_{a,b}$ ratio of $\frac{C}{A_{a,b}}=0.212$. \citet{dw} extended these calculations to multi-level transitions in CO, CS, and SiO using molecular data from \citet{green_chapman_78} and \citet{robinson_81}. \citet{cortes} moved further to include the influence of anisotropic radiation, a mixture of 1-D and 2-D LVG geometries, and updated molecular data from \citet{flower2001} motivated by their own observations and those of \citet{lai03}. \citet{yl10} developed an independent G-K numerical model incorporating these prior physical approaches, but excluding anisotropic radiation, and similarly focused on star formation. The code uses updated molecular data from the Leiden Atomic and Molecular Database (LAMDA) \citep{lamda_database}. For technical details concerning the underlying algorithm implementation, we refer the reader to \citet{yl10}. In the current work the \citet{yl10} G-K code was applied in a CSE environment and a source of radiative anisotropy was introduced to model emission from the central AGB star; this is discussed further in section \ref{sec:pol_discussion}. Representative densities and temperatures for the CSE, as tabulated in Table \ref{tab:t_CSEpara}, differ from those in star-forming regions described above. The model used here considers an uniform magnetic field and uniform velocity gradient through the slab of material with constant thickness and density, as a first-order approximation. The code supports the following three types of velocity gradient geometry (as depicted in Fig. \ref{fig:vgradient_geo}: i) 1-D velocity anisotropy parallel to the magnetic field; ii) a 2-D velocity gradient along the 2-D plane perpendicular to the magnetic field; and, iii) a mixed model of (i) and (ii), yielding a cone-shaped velocity gradient lobe. We used model (iii) to model bipolar outflows. 

\boldred{We note important new theoretical work by \citet{lankhaar_2020}.}
\subsubsection{Observed linear polarization and the G-K model predictions} 
\label{sec:pol_discussion}
\begin{figure}
	\center
	\includegraphics[scale=0.8]{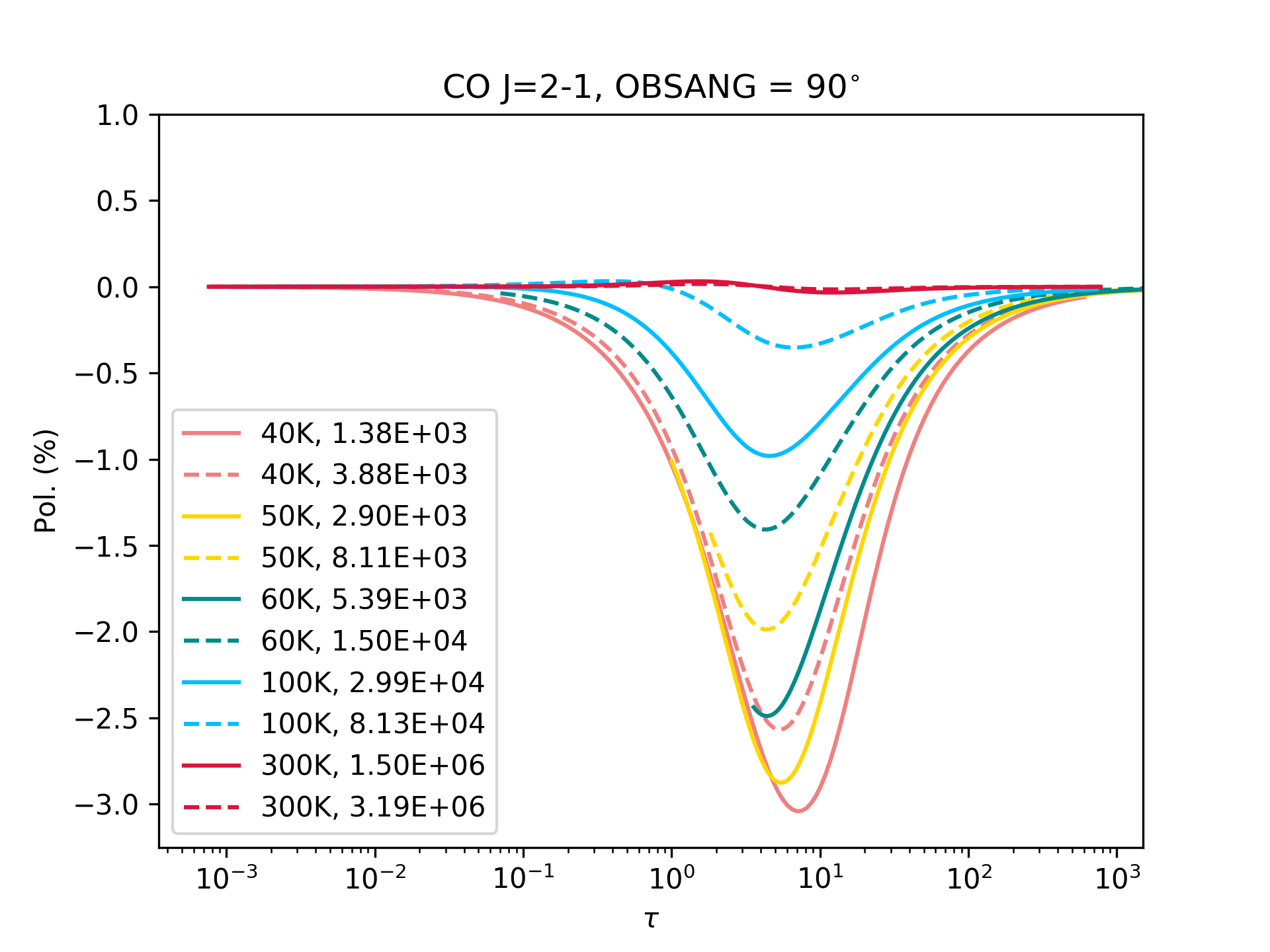}
	\caption{The predicted polarization signal level from our G-K modeling using mix model (corresponding to $\alpha=0.1$ as discussed in \citet{cortes}), density and temperature profile calculated based on R Leo case as listed in Table \ref{tab:t_CSEpara}. The colors denote the temperature from low to high. The styles denote the lower bound (solid) and the upper bound (dashed) of the density profile listed in Table \ref{tab:t_CSEpara}. }
	\label{fig:COpol_model}
\end{figure}

\begin{figure}
	\center
	\includegraphics[scale=0.8]{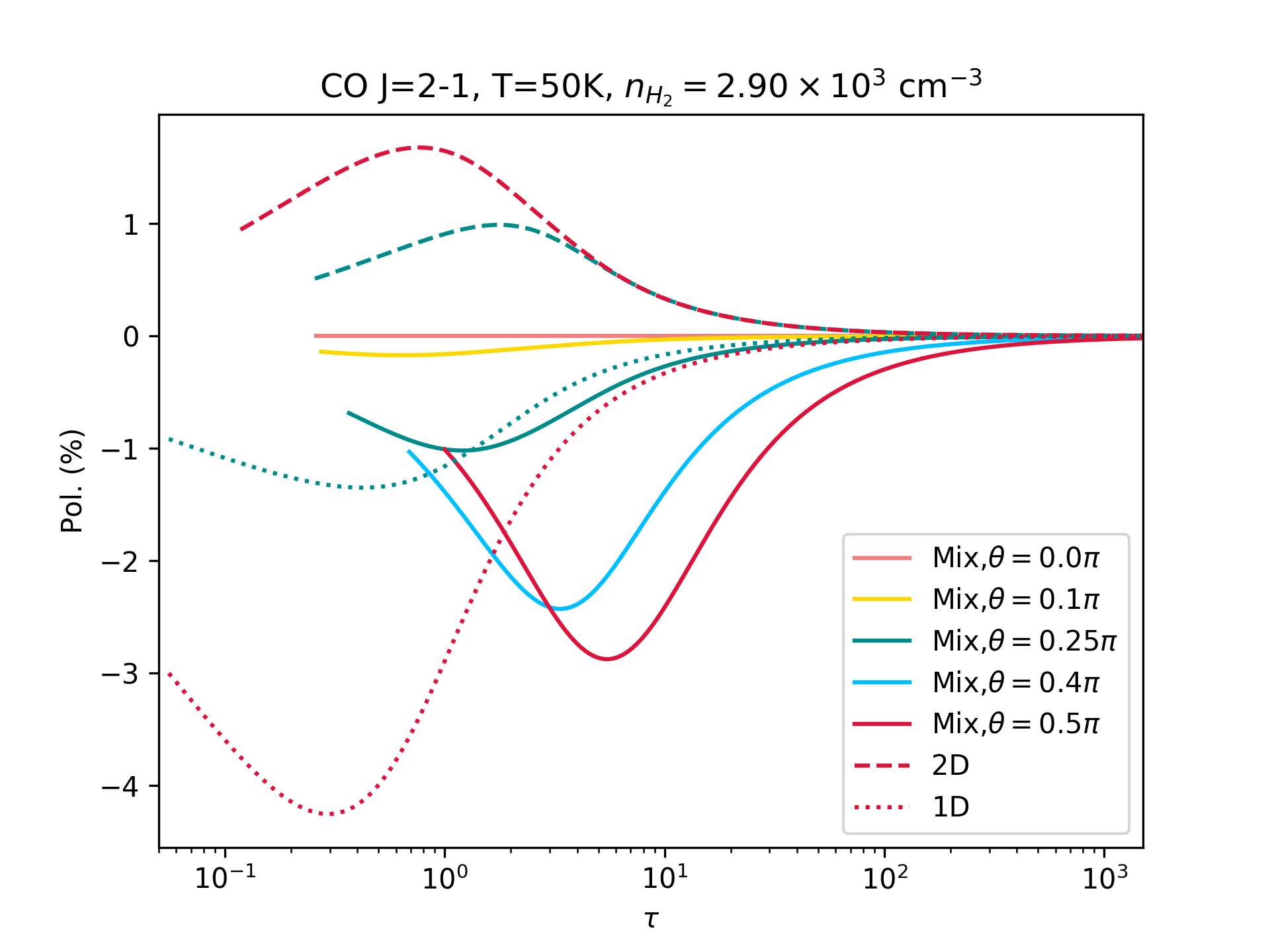}
	\caption{A comparison of the predicted G-K polarization profiles for the $J=2-1$ CO line under various choices of velocity gradient geometries (Figure \ref{fig:vgradient_geo}) and the line-of-sight (LOS) viewing angles. The angle $\theta$ (see color legend) refers to the angle between the LOS direction and the presumed magnetic field direction. Line style indicates the 1-D, 2-D, and Mix LVG geometry models. The CSE parameters used here are: T = $50$K, n(r) = $2.9\times 10^{3}$cm$^{-3}$. }
	\label{fig:COpol_model_LVGs}
\end{figure}

The predicted G-K fractional linear polarization is shown as a function of optical depth, temperature, and density in Fig.\ref{fig:COpol_model} for the CSE model described above and summarized in Table \ref{tab:t_CSEpara}. A mixed LVG velocity gradient geometry (Fig. \ref{fig:COpol_model_LVGs}(iii)) was adopted in the G-K modeling as this geometry is closest to the modest axisymmetry possible in the late AGB phase. The effect of varying the velocity gradient geometry for a fixed representative temperature and density is shown in Fig. 
The legend in Fig. \ref{fig:COpol_model} indicates temperature (line color) and density (line style) thereby representing various effective distances from the central star. The densities are upper and lower bounds in Table \ref{tab:t_CSEpara} for $\beta \in \{0,5\}$. For G-K emission, the sign $m_l < 0$ implies that the EVPA is parallel to the magnetic field. If the linear polarization in Fig. \ref{fig:carmaco} is due to the G-K effect then the EVPA and magnetic field are aligned in the Figure \ref{fig:COpol_model_LVGs}.

Our G-K model predicts $m_l \sim 3\%$ (Fig. \ref{fig:COpol_model}) which is consistent with our observational result for the CO emission toward R Crt. However, the fractional linear polarization $m_l \sim 9.7\%$ detected in CO toward R Leo exceeds our G-K modeling predictions. We note however that if the polarized emission originates from a region more compact than the associated Stokes $I$ emission then interferometric spatial filtering (leading to missing flux density) may overestimate $m_l$. 

An anisotropic radiation field can also contribute to linear polarization, reaching a maximum of order $m_l \sim 5-10\%$ \citep{morris1985,cortes}. To explore this contribution we added an anisotropic radiation source in our G-K model. We considered the relative radio brightness of the central star and a dust clump at the dust-formation radius. The central star is assumed to be of order several thousand Kelvin and to subtend an angle $\sim 0.01$ rad from the circumstellar CO region at a radius $r\sim 200R_*$. Under these parameters the predicted G-K linear polarization peaks at the same optical depth as in Fig. \ref{fig:COpol_model} but $m_l$ is not significantly increased. Over a broader parameter range of physical, and even unphysical, conditions we find that a fractional linear polarization exceeding $m_l > 5\%$ cannot readily be generated by our G-K model including anisotropic radiation.

\boldred{We do not expect the fractional G-K linear polarization $m_l$ to be identical between the sources due to differing local conditions (Fig. \ref{fig:carmaco}). \citep{vlemmings12} report a value $m_l\sim 13\%$ for IK Tau in this CO transition although this value is similarly believed elevated due to interferometric flux density filtering in Stokes $I$.}

Lastly, the uncertainty in $T_*$ noted earlier translates to uncertainty in the CSE temperature following Equation \ref{eq:temp} but does not have a significant effect on our inferred G-K linear polarization magnitude.
%\AKnote{Need a more detailed caption for Figure 7. Need to define all symbols and abbreviations and describe the nature of the geometry depicted in each panel.}
%\AKnote{also need for detailed caption for Figure 8}
\\
\subsection{SiO maser polarization}
We detect linear polarization in the $v=1,\ J=5-4$ SiO maser emission toward R Crt at $m_l \sim 17.2\%$ and R Leo at $m_l\sim 34.6\%$ (Table \ref{tab:table3}). At the closest matching velocities in the $v=1,\ J=2-1$ SiO maser spectra presented by \citet{herpin06} $m_l \sim 15\%$ for R Crt and $m_l \sim 26\%$ for R Leo. In the same $v=1,\ J=2-1$ SiO maser transition, \citet{wiesemeyer09} report a fractional linear polarization toward R Leo of $m_l \sim 24-26\%$. We note however that these are different SiO rotational transitions at differing spatial resolution and that SiO maser emission is highly time-variable \citep{pardo04}. Nonetheless these results are broadly consistent with the current work.
%Concerning the same maser transition with our work, $v=1,\ J=5-4$ SiO line, high fractional linear polarization has been observed on other evolved stars, of even above $50\%$ level \citep{shinnaga2004,vlemmings2011}. 
Linear polarization in the current $v=1,\ J=5-4$ SiO maser transition has been observed toward other target sources including VY CMa at $m_l \sim 10-60\%$ \citep{shinnaga2004} and VX Sgr at $<m_l>=26\pm16\%$ \citep{vlemmings2011}. \citet{watson09} argues that high fractional linear polarization in circumstellar SiO maser emission (approaching $\sim 50\%$) would require an unphysical degree of maser saturation; higher values of $m_l$ are more readily explained by anisotropic pumping. High fractional linear polarization in high$-J$ SiO maser transitions strengthens the case for anisotropic pumping \citep{lankhaar2019}.

We detect peak fractional circular polarization $m_c \sim 5.7\%$ in the $v=1, J=5-4$ SiO maser line toward R Leo (Table \ref{tab:table3}), but do not detect statistically significant circular polarization in this transition toward R Crt perhaps due to sensitivity. For reference, \citet{herpin06} report $m_c \sim 9-10\%$ toward R Leo in the $v=1, J=2-1$ SiO maser transition.

\subsection{Comparison with other intrinsic alignments}

\begin{figure}
  \centering
  \begin{tabular}[b]{@{}p{0.45\textwidth}@{}}
    \centering\includegraphics[width=1.0\linewidth]{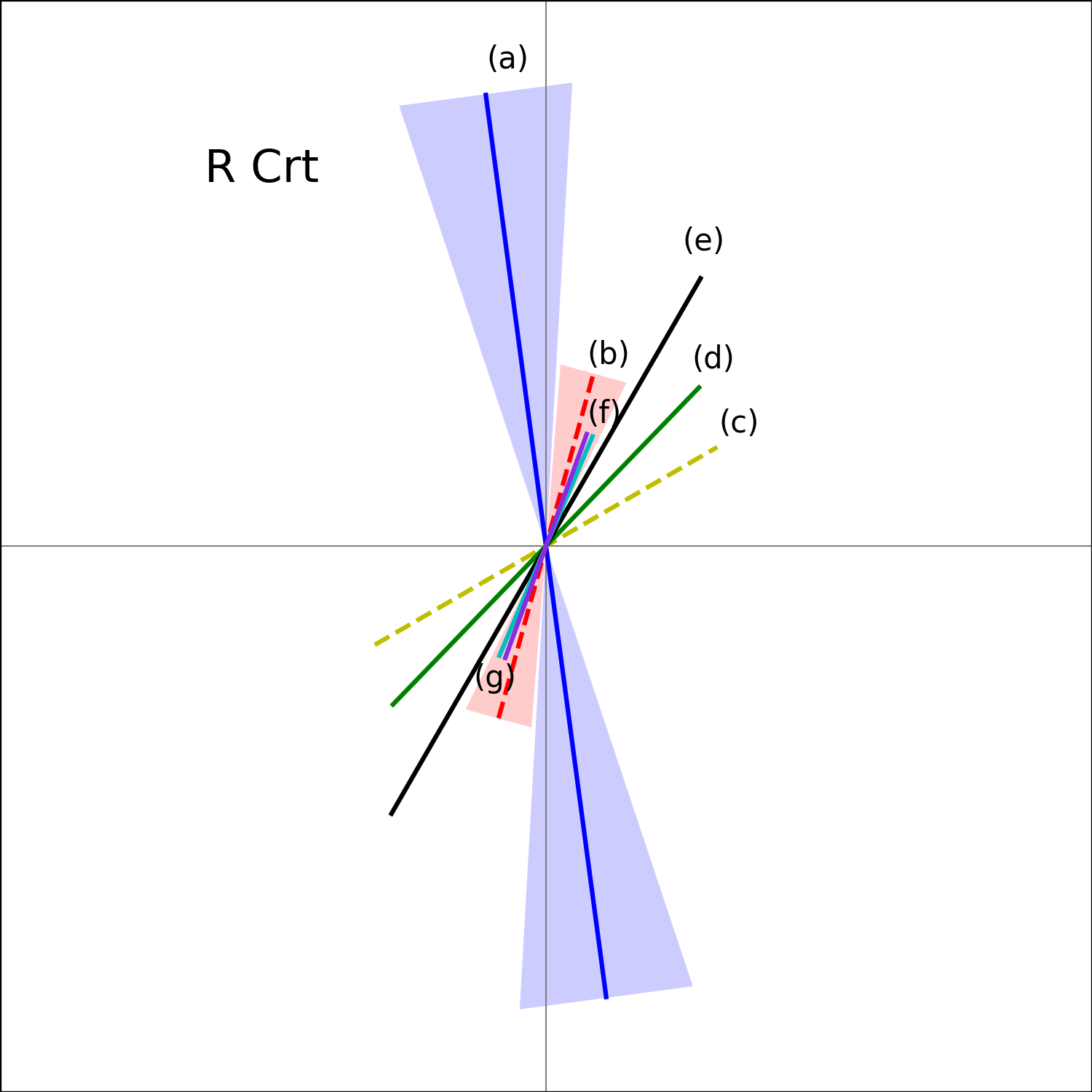} \\
    \centering\small (a) R Crt
  \end{tabular}%
  \quad
  \begin{tabular}[b]{@{}p{0.45\textwidth}@{}}
    \centering\includegraphics[width=1.0\linewidth]{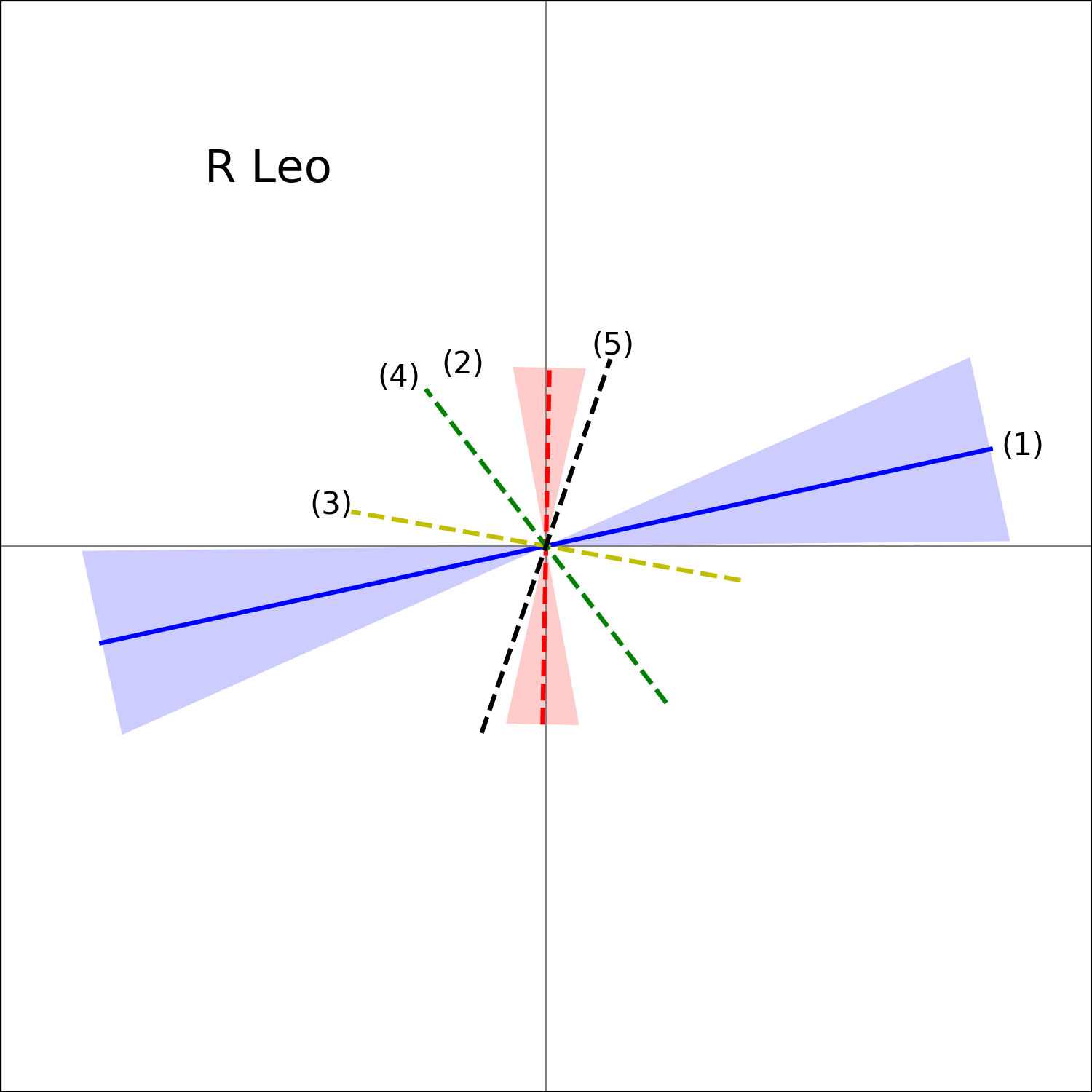} \\
    \centering\small (b) R Leo
  \end{tabular}
  \caption{Schematic plot from the compiled results. The labeling refers to the literature cited in Table \ref{tab:table_compile}}
  \label{fig:compile_2}
\end{figure}

In this section we consider the relation between the magnetic field orientation measured in the current work for R Crt and R Leo and other intrinsic alignments published for these sources in the literature.
The compiled results are enumerated and described in Table \ref{tab:table_compile}. This table lists the intrinsic position angle (N through E), the observations (and telescope or instrument) from which this angle was inferred, and the corresponding citation.
The compiled results are plotted schematically in Fig. \ref{fig:compile_2} where the shaded opening angle indicates total uncertainty in orientation $\sigma_{tot} \sim \sqrt{\sigma_{n}^{2}+\sigma_{s}^{2}}$, when available.

\begin{table}[ht!]
  \centering
  \caption{Intrinsic position angles and alignments measured for R Crt and R Leo.}
  \label{tab:table_compile}
  \begin{tabular}{cllll|c}
  \hline
    Source & Position angle        & Emission & Telescope or Instrument & Reference & Label\\
    \hline
    \hline
    R Crt  & $\phi_{B} = 7.6^{\circ}\pm 1.0^{\circ}\pm 10.8^{\circ}$  & CO $(J=2-1)$         & CARMA     & This work & (a)\\
    {}     & $\chi=-15.4^{\circ}\pm 0.7^{\circ}\pm 10.8^{\circ}$     & SiO $(J=5-4, v=1)$   & CARMA     & This work  & (b)\\
    {}     & $\chi=120^{\circ}/-60^{\circ}$          & SiO $(J=2-1, v=1)$    & IRAM 30 m & \citep{herpin06} & (c)\\
    {}     & $\phi_{outflow}=136^{\circ}/-44^{\circ}$ & H$2$O maser          & Japanese VLBI Network & \citep{ishitsuka01} & (d)\\
    {}     & $\phi_{B}=-30^{\circ}$                   & OH 1667-MHz maser    & MERLIN array & \citep{szymczak99} &(e)\\
    {}     & $\phi_{star}=157^{\circ}$                   & Mid-IR (Stellar elongation)   & VLTI/MIDI & \citep{paladini2017} &(f)\\
    {}     & $\phi_{outflow}=160^{\circ}$ & Visible (Dust scattering) &  VLT/SPHERE/ZIMPOL  & \citep{khouri_etal2020} & (g)\\
    \hline
    R Leo  & $\phi_{B} = -77.7^{\circ}\pm 1.7^{\circ}\pm 11.6^{\circ}$& CO $(J=2-1)$         & CARMA     & This work & (1)\\
    {}     & $\chi=-1.1^{\circ}\pm 0.03^{\circ}\pm 11.6^{\circ}$     & SiO $(J=5-4, v=1)$   & CARMA     & This work & (2)\\
    {}     & $\chi=80^{\circ}$                        & SiO $(J=2-1, v=1)$   & IRAM 30 m & \citep{herpin06} & (3)\\
    {}     & $\chi=[-40^{\circ},-35^{\circ}]$ \tablenotemark{a}        & SiO $(J=2-1, v=1)$   & IRAM 30 m & \citep{wiesemeyer09} & (4)\\
    {}     & $\chi=-19^{\circ}\pm 65^{\circ}$         & SiO $(J=2-1, v=1)$   &NRAO 12 m Telescope & \citep{glenn_etal_03} & (5)\\
    \hline
  \end{tabular}
  \tablenotetext{a}{The listed EVPA range is extracted from their Fig.2; the geometric mean of this range is plotted in Figure \ref{fig:compile_2}. }
  \tablecomments{In this table, $\chi$ denotes an SiO emission EVPA, $\phi_B$ an inferred G-K magnetic field direction, $\phi_{outflow}$ an outflow direction, and $\phi_{star}$ the position angle of a stellar elongation.}
\end{table}

Several caveats apply when interpreting these alignment and orientation data.  
The measured SiO maser EVPA has a $90^{\circ}$ ambiguity ($\parallel, \perp$) with respect to the magnetic field, depending on the angle between the line of sight and the magnetic field line relative to the Van Vleck angle $\theta_{F} \sim 55^{\circ}$ \citep{gkk73}. SiO EVPA orientations are plotted accordingly with dashed lines in Fig. \ref{fig:compile_2} to highlight this ambiguity.
The physical conditions and associated radial distances sampled by the SiO, H$_{2}$O, OH, and CO molecules differ significantly as noted earlier. The length of each line is monotonically proportional to the the distance scales traced by each transition, but not drawn strictly to scale for clarity of presentation.   
Fig. \ref{fig:compile_2} shows that for R Crt, the measured angles are more narrowly confined compared to the case of R Leo.

%\AKnote{Fig. 10 needs a more detailed caption; can refer to the text for more details but needs a more self-contained description in caption}
%In the following plot, our results and the results from the literature is compiled for further examination. 
%%MMM reasons...
%To name a few, it is expected in the difference of the location of the CO (intermediate CSE, $\gg$10 R$_{*}$) and the used maser sources ($<$10 R$_{*}$) can be found.
%This can mean that the larger scale field lines ...
%%
% Point is get to brief talking about each source features. 
\section{Conclusions}
We have detected linear polarization in thermal $J=2-1$ CO line emission from the CSE of the TP-AGB stars R Leo and R Crt in $\lambda 1.3$ mm full-Stokes observations with CARMA.
The fractional linear polarization measured in the $J=2-1$ CO line for R Crt ($m_l \sim 3.1\%$) is consistent with the predicted modeled G-K signal strength, while for R Leo the measured fractional linear polarizaion $(m_l \sim 9.7\%)$ is higher than expected. Our G-K modeling is able to resolve the directional degeneracy, suggesting that the measured EVPA parallel to the magnetic field lines. 
We also have detected linear polarization in the $v=1, J=5-4$ SiO maser line toward the CSE of R Crt and R Leo in the range $m_l \sim 17-34\%$. 
We detect circular polarization in this maser line toward R Leo but not toward R Crt, the latter possibly due to sensitivity limitations. 
We detect continuum emission toward R Crt and R Leo in the wide-band spectral windows. However, no successful detection of polarized continuum emission is achieved likely due to sensitivity limitations and the relative weakness of dust emission to the stellar continuum.
These results with CARMA from both observation and modeling demonstrate that the G-K effect is a viable means of extracting information concerning magnetic field morphology in the CSE around late-type evolved stars at various depths in the envelope. These observations are profoundly sensitivity-limited compared to modern telescopes such as ALMA but are the first such observations with CARMA and confirm the scientific importance of observations of this nature \citep{vlemmings12}.
%\AKnoteM{include vlemmings reference to IKTau GK paper}.
At greater sensitivity, future in-depth morphological mapping of the circumstellar magnetic field around late-type stars is possible using a range of molecular species and the associated transitions. 

\appendix

\section{Polarization and calibration quality assurance}\label{sec:qa}
This Appendix described the polarization quality assurance tests applied to the fully calibrated data.

The amplitude of the derived polarization leakage terms $|\mathbf{D_m}|$, across all spectral windows and both observing runs, ranged from few percent to a maximum of $\sim 20\%$, consistent with results from prior CARMA continuum polarimetry \citep{hullplambeck}. Although the $\mathbf{D_m}$ are known to depend on time and frequency, a moderate level of instrumental stability is expected. 
We therefore examined the rank correlation of the antenna-based $|\mathbf{D_m}|$ against those obtained from a separately-reduced CARMA continuum polarimetry archival data set\footnote{CARMA project ID: c1217; Observing Block ID: c1217.2D\_2303c279.2} \citep{agnSurvey_bower_2014}. The Spearman rank-order correlation \citep{zwillinger_kokoska_2000}, Kendall's tau \citep{knight1966}, and the Pearson correlation coefficient \citep{pearson1895} were computed using SciPy\footnote{http://www.scipy.org} for both the R Crt and R Leo data, across all spectral windows, and against the values of $|D^R_m|$ and $|D^L_m|$ derived from the archival data. The correlation coefficients fell within the range $r \in [0.1 \sim 0.6]$ with a maximum of 15 antennas in the current data. A median value $r \sim 0.4$ was obtained, indicating moderate correlation.

In addition, the cross-polarized visibility data showed significantly less scatter when plotted in the Re-Im plane before and after polarization calibration. Similarly the statistical distribution of the R-L phase difference $\phi_{RL,m}$ measured from the visibility data was narrower after full polarization calibration. Further, the dynamic range of the calibrator polarized intensity maps improved from $\sim 30$ to $\sim 150$ after polarization calibration.

Underlying phase calibration accuracy was assessed from the phase structure function of the calibrated parallel-hand visibility data in the continuum windows for the calibrators, averaged over frequency. For R Crt, the residual phase rms obtained is $\sigma_{\phi} \sim 9.8^{\circ}$ and for R Leo $\sigma_{\phi} \sim 17.0^{\circ}$. Both phase structure functions showed a slow increase of $\sim 10^{\circ}$ over a $uv$ distance from $5.0$ k$\lambda$ to $40.0$ k$\lambda$. The compactness of the residual phase distribution however is evidence of sufficiently robust phase calibration. 

\citet{hullplambeck} list the following sources of systematic error in polarization observations with CARMA: i) the need to de-bias linear polarization measurements \citep{wardle_kronberg_74}; we apply this correction as noted above; 
ii) limitations due to polarization leakage uncertainties; 
iii) absolute EVPA determination accuracy; and,
iv) direction-dependent beam polarization (beam squint and beam squash). 

\citet{hullplambeck} note that EVPA interpretation requires caution for weakly linearly polarized sources ($m_l \leq 0.5\%$).
The current data exceed this $m_l$ threshold. However, to verify the accuracy of the absolute EVPA aligment (or equivalently the residual uncorrected R-L phase difference at the reference antenna), we compared the measured EVPA values of the calibrators (3C279, OJ287) in our data against surveys or other contemporaneous observations.

Our current data yielded an error-weighted average EVPA measurement for 3C279 of $36.6^{\circ}\pm 0.07^{\circ}$ and $33.9^{\circ}\pm 0.08^{\circ}$, and for OJ287 of $-16.5^{\circ}\pm 0.33^{\circ}$ and $-19.0^{\circ}\pm 0.40^{\circ}$, for the LSB and USB continuum spectral windows respectively. As noted above these 500 MHz spectral windows are centered on $\sim 215$ GHz and $\sim 230$ GHz respectively.
Applicable external EVPA values were obtained from the MOJAVE project \citep[][private communication]{majove}, the VLBA-BU-BLAZAR program \citep[][private communication]{jorstad16,jorstad17}, and the POLAMI project\footnote{http://polami.iaa.es} \citep{agudo_2018,agudo_2018a}. 
We selected external measurements falling within $\pm 14$ days from our observations. Available observations were limited and we considered data within a frequency range of $43.0 \sim  230.5$ GHz.
The resulting 14-day averaged EVPA for 3C279 thus obtained is $42.88^{\circ} \pm 10.83^{\circ}$; and for OJ287 is $159.85^{\circ} \pm 11.58^{\circ}$. As a global reference, we note that when averaging over 1000 days the EVPA value obtained for 3C279 is $135.4^{\circ} \pm 18.3^{\circ}$ (MOJAVE) and $151.4^{\circ} \pm 26.0^{\circ}$ (VLBA-BU-BLAZAR); for OJ287 this global average over 1000 days is $21.5^{\circ}\pm 4.4^{\circ} $ (MOJAVE) and $29.3^{\circ}\pm 17.7^{\circ} $ (VLBA-BU-BLAZAR).
In summary, our measured EVPA of both OJ287 and 3279 were found to be consistent with the external contemporaneous EVPA measurements in broadly comparable frequency bands. 

For (iv), beam squint is embodied in a double-lobed pattern in Stokes V maps, while beam squash is evident as a cloverleaf pattern in the Stokes Q and U maps \citep{hullplambeck}. At $\lambda \sim 1.3$mm wavelength, the FWHM of the primary beam for the CARMA 10-m antennas is $\sim 30$", and for the 6-m antennas it is $\sim 56$" \citep{hullplambeck}. Our source extent is $<$ 10" and on-axis and therefore well within the inner primary beam. Accordingly we do not expect strong primary beam polarization effects. We find no evidence of a quadrupolar squash pattern in the Stokes $Q$ or $U$ images in the current data. We expect our quasar calibrators to have low intrinsic circular polarization \citep{RL1979_radiativebook}. The residual Stokes $V$ signal in our calibrators is $m_c \sim 0.32{\%}$ for 3C279 (in the R Crt observing run), and $m_c \sim 0.77{\%}$ for OJ287 (in the R Leo observing run) after full calibration. \boldred{To zeroth-order we assume $\sigma_{m_c} \sim \sigma_{m_l}$ and ${\bar m_c} \sim 0$ and conservatively estimate an approximate intrinsic linear polarization accuracy $\sigma_{m_l}\sim 0.3-0.8\%$ accordingly. In practice the calibrators have low, but non-zero circular polarization \citep{agudo_2018}, amongst other factors and this is an approximate estimate only.}

\section{CSE parameters}
\label{sec:cse}
Detailed modeling of the kinematics, dynamics, and chemical structure in the CSE requires numerical approaches \citep{bowen88,humphreys_1996,ireland_2008,ireland_2011} given the inherent complexity of pulsation shock structure, mass loss, and chemistry in this region. However, the G-K modeling in the current work allows use of input profiles from prior semi-analytic models of the CSE, as described further in this Appendix.

The G-K effect allows study of the AGB CSE at different depths due to molecular stratification by chemical abundance and radiative excitation effects. The mean CO distribution extends broadly from the photospheric radius $(\sim 1R_{*})$ to the outer circumstellar envelope $(\gtrsim 10^{4} R_{*})$ \citep{aarv2018_hofner}, with an assumed constant mean fractional abundance $X_{CO} = \left(\frac{n_{CO}}{n_{H_{2}}}\right)$ \citep{duari1999,decin10} until the outer photo-dissociation radius; this encompasses a diverse range of local conditions however. In contrast, SiO is prone to depletion in the inner CSE including due to adsorption onto dust grains \citep{teyssier06,decin10,chem_text,gobrecht16}. \boldred{In W Hya the SiO e-folding radius is $r_e \sim 85R_*$ \citep{gonzalez03,khouri14}, with dissipation at radius $r \sim 200R_*$ \citep{khouri14}. Modeling of IK Tau by \citet{decin10} shows a depletion of SiO by a factor $\sim 40$ at $r \sim 180R_*$, with an outer photo-dissociation accordingly well within CO.}

\subsection{Temperature profile}
%(Either the power law or from the differential equation)
\paragraph{Power-law temperature profile} Following \citet{bowen88} as cited by \citet{cherchneff92}, the kinetic temperature dependence on radius is approximated as a power-law:
\begin{equation}
\frac{T(r)}{T(R_{*})} = \left( \frac{r}{R_{*}} \right)^{-\alpha}
\label{eq:temp}
\end{equation}
where the exponent index, $\alpha$, is often adopted to be $0.6$ \citep{cherchneff92,decin10,gobrecht16}. The functional form of this relation can be derived from the assumption of \boldred{radiative equilibrium} and a gray stellar atmosphere \citep{bowen88}.
For R Leo the effective stellar blackbody temperature $T_{*}$ is estimated from modeling of the spectral energy distribution (SED) to be $T_{*} = 1.8 \times 10^3K$ \citep{schoier13,success15}. We adopt an inner dust condensation radius $R_{c} \sim 1.6 \times 10^{14}\ \rm{cm}$ from SED modeling \citep{success15}. These parameters predict a thermal temperature at the dust-condensation radius $T(R_c) \sim 5.3 \times 10^2K$. 

\subsection{Density profile}
Semi-analytic solutions exist for approximate conditions within broad radial zones in the CSE. We adopt an inner region, extending from the photosphere to the dust formation radius $R_c$, and an intermediate-outer CSE region extending beyond $R_c$ to the outer edge of the photodissociation zone, and imposing continuity in the density relation.

\paragraph{I. Inner CSE}
The inner CSE includes the shock formation radius $R_0$ \citep{hill_1979}. We adopt the functional form of the radial density profile $n(r)$ in the shock-extended region derived by 
\citet[eq.17]{cherchneff92} by considering averaged physical quantities in the momentum equation. Further, \citet{cherchneff92} derive an expression for $R_0$ using numerical model results from \citet{willson_1986}. This expression supports the approximation $R_0 \sim R_*$, which we adopt here.

\paragraph{II. Intermediate-outer CSE}
In this region we assume the velocity $v_{\beta}(r)$ is governed by the classical $\beta$-law \citep{LamersCassinelli99,success15}; this yields a velocity gradient:
\begin{equation}
\frac{dv_{\beta}}{dr} = \frac{\beta}{r} \left[ v_{c} + (v_{\infty} - v_{c}) \left( 1-\frac{R_{c}}{r} \right)^{\beta -1} - v_{c} - (v_{\infty} - v_{c}) \left( 1-\frac{R_{c}}{r} \right)^{\beta} \right]
\equiv \frac{\beta}{r} [v_{\beta -1} - v_{\beta}]
\label{eq:betalaw}
\end{equation}
where $v_{c}$ is the expansion velocity at $R_{c}$ and $v_{\infty}$ is the terminal expansion velocity. Typically $v_c \sim 3$ km/s based on the approximate sound speed at this radius \citep{success15}. We assume $v_{\infty} \sim {\rm const.}$, with $v_{\infty} \sim 8.5$ km/s for R Leo and $v_{\infty} \sim 12.0$ km/s for R Crt in CO emission \citep{success15}. We adopt a range in velocity exponent index $\beta = 0 \sim 5$ \citep{success15}; 
where $\beta = 0$ implies constant isotropic outflow. 

Continuity in the equation of mass yields,
\begin{equation}
\frac{d \rho}{d r} + \frac{\rho}{r} \left[ (2-\beta) + \beta \cdot \frac{v_{\beta -1}}{v_{\beta}} \right] = 0
\label{eq:ode_tosolve}
\end{equation}
where the quasi-static state and zeroth-order approximation $\vec{v} = v \hat{r}$ is assumed.

We used the SciPy\footnote{https://www.scipy.org} package $\mathtt{odeint}$ to solve the differential equation in Equation \ref{eq:ode_tosolve} for $n(r)$ imposing continuity across the inner-region boundary at $r=R_c$. We model the CSE of R Leo in this manner, adopting parameters: $R_*=1.4\ \rm{AU}$ \citep{debeck10}, $T_*=1800\ \rm{K}$, \citep{schoier13}, $R_c=10.7\ \rm{AU}$ \citep{success15}, $v_{\infty}=8.5\ \rm{km/s}$ \citep{success15}, and ${\dot M}=1.1\times10^{-7} M_{\odot}/{\rm yr}$ \citep{success15}. Photospheric temperature estimates in the literature fall in the ranges $T_* \sim 1800-2100\ {\rm K}$ \citep{schoier13,ramstedt14,gonzalez03} and $T_* \sim 2850-2890\ {\rm K}$ \citep{debeck10,vlemmings_etal19}. 

The boundary condition continuity was enforced by estimating the number density at the dust condensation radius as $n(R_c)=\frac{\dot M}{4\pi R_{c}^2v_{c}}$, assuming a photospheric density $10^{14}\ \rm{cm}^{-3}$ \citep{gobrecht16}, and adopting the functional form of the inner CSE density profile \citep{cherchneff92} discussed above.

\begin{figure}
  \centering
  \begin{tabular}[b]{@{}p{0.7\textwidth}@{}}
    \centering\includegraphics[width=1.0\linewidth]{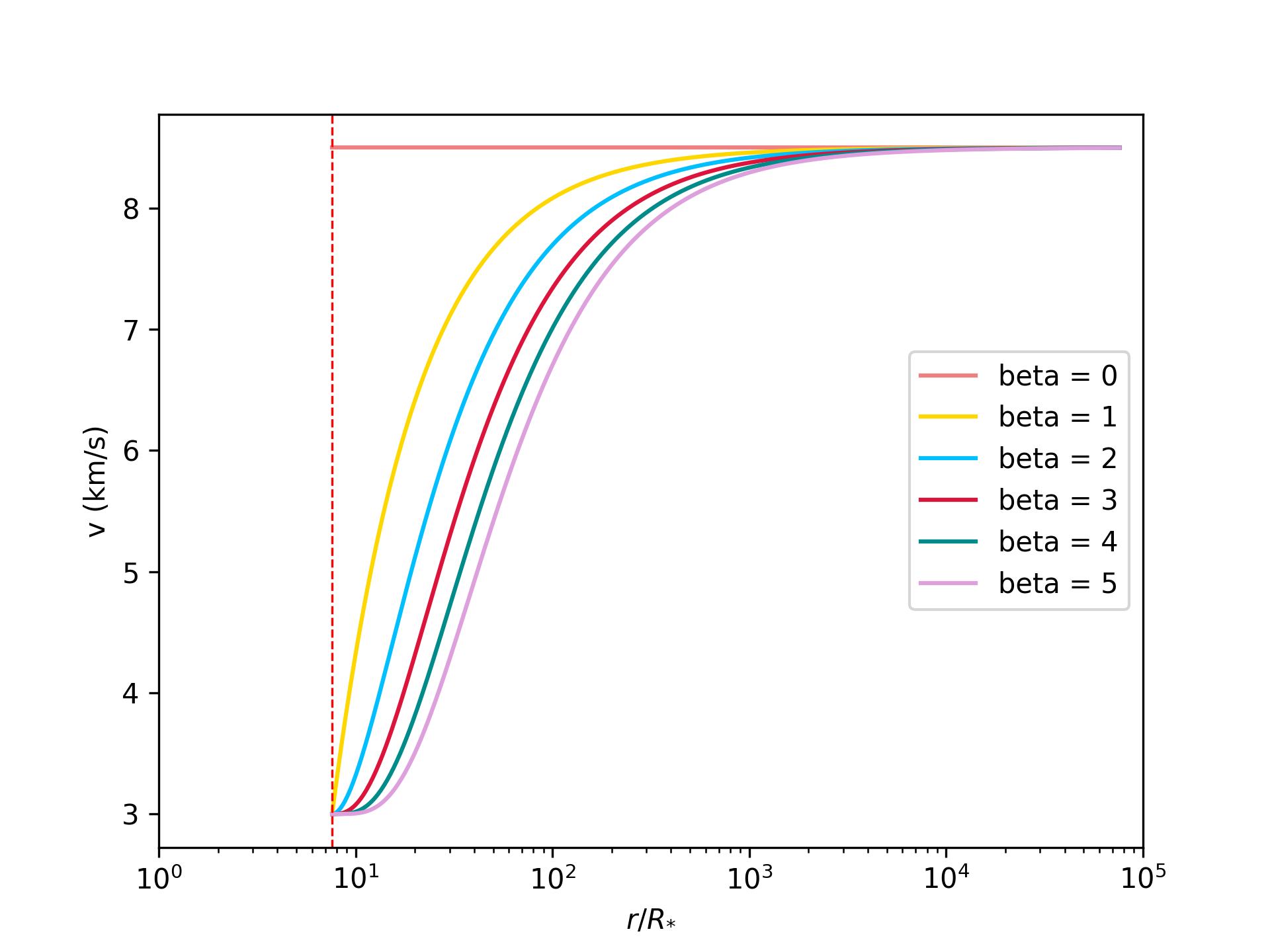} \\
    \centering\small (a) Velocity profile for the range $\beta = 0\sim5$ and based on the adopted physical parameters of R Leo. 
  \end{tabular}%
  \quad
  \begin{tabular}[b]{@{}p{0.7\textwidth}@{}}
    \centering\includegraphics[width=1.0\linewidth]{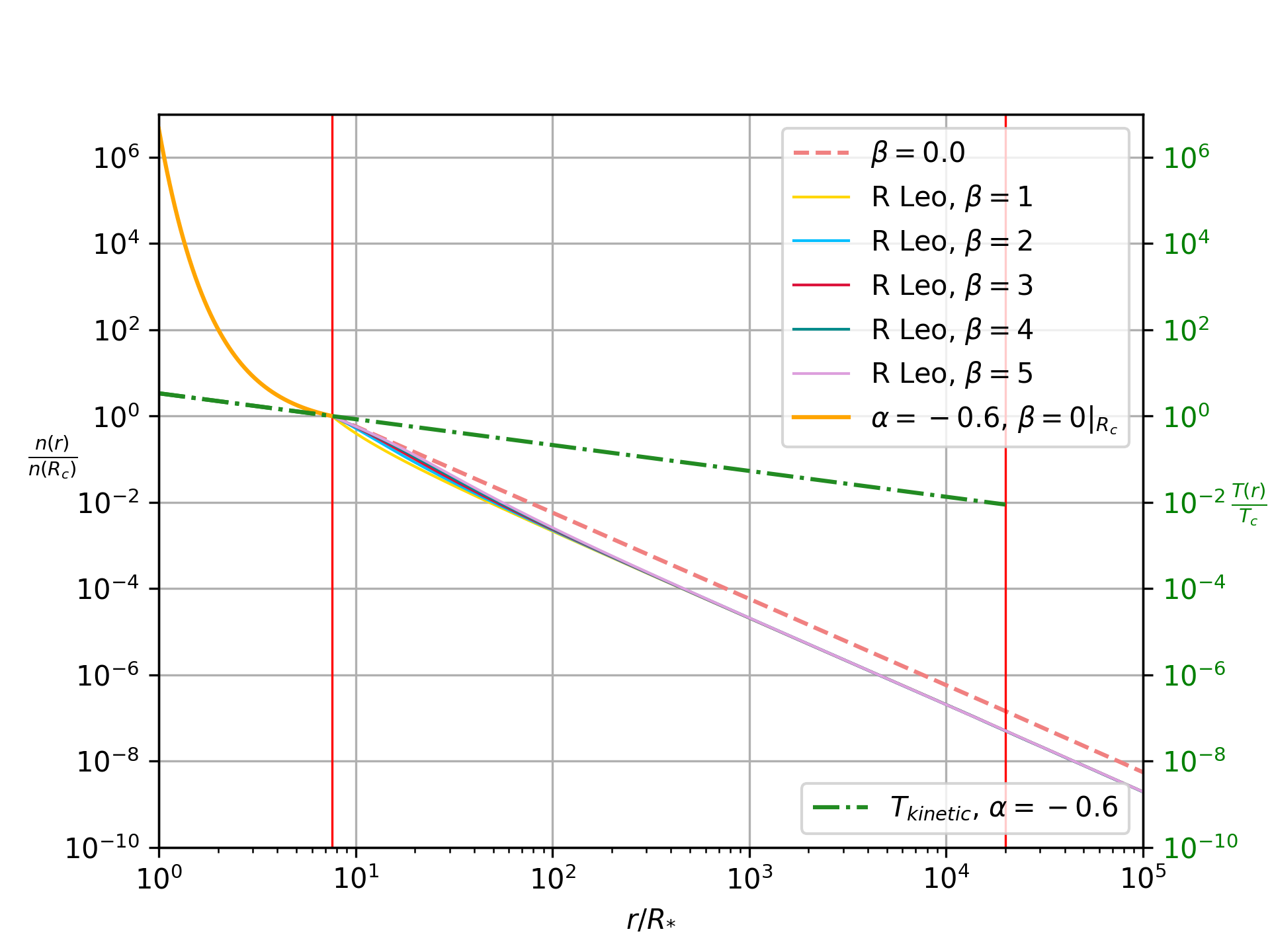} \\
    \centering\small (b) Density and temperature profile based on the listed physical parameters of R Leo. The left vertical axis labels the normalized number density, $ \frac{n(r)}{n(R_{c})}$. The vertical axis on the right marks the scale for the normalized temperature profile $\frac{T(r)}{T_{c}}$. The green dash-dotted line portrays the calculated temperature profile. The red dashed line is the density profile for a model with $\beta=0$ and $v_{\infty}=3.0\ {\rm km/s}$. The remaining density profile lines overlap significantly at larger $\frac{r}{R_*}$ and include models with $\beta \neq 0, v_c=3.0\ {\rm km/s},v_{\infty}=8.5\ {\rm km/s}$. 
  \end{tabular}
  \caption{CSE parameters as GK modeling input. }
  \label{fig:CSE_para_figs}
\end{figure}

The derived profiles in velocity, density, and temperature for R Leo are displayed in Fig. \ref{fig:CSE_para_figs}; they are developed as required input to the G-K modeling only. The derived CSE parameter sets (temperature $T(r)$ and H$_{2}$ number density $n(r)$ profiles) are summarized in Table \ref{tab:t_CSEpara}. For the G-K modeling we assume that we lie within the CO photo-dissociation radius. In the G-K modeling, the general H$_2$ density defines the collisional partner for CO but the modeling is not highly sensitive to the exact CO abundance ratio assumed. 

The semi-analytic CSE models do not include spectral line excitation or emission. However, the adopted mass-loss rate and CO outflow kinematics should be consistent with the predicted CO emission intensity.  Recent modeling initiatives in the literature to estimate the mass loss rate ${\dot M}$ for R Leo from CO emission parameters are enumerated in Table \ref{tab:CO_RLeo}; \boldred{we refer the reader to these papers for further details of each model}. The parameters are as defined above. We have applied our CSE model to each set $k$ of parameters in Table \ref{tab:CO_RLeo} to derive predicted density profiles $n_k(r)$. The maxima and minima of $n_k(r)$ over $k$ are listed in Table  \ref{tab:CO_RLeo_n} at the radii sampled in Table \ref{tab:t_CSEpara}; these density ranges are broadly consistent. We therefore believe that the density profiles predicted by our CSE model are conservative but reasonable estimates within the uncertainty in ${\dot M}$.

\begin{table}[ht!]
  \centering
  \caption{The radial profiles in temperature $T(r)$, density $n(r)$, and velocity (as represented by parameter $\beta$ in Equation \ref{eq:betalaw}), generated from the CSE analytic models. \label{tab:t_CSEpara}}
  \begin{tabular}{cccccc}
  \hline
  \hline
   Source & r  & T & n(r)$^{a}$ & $\beta$   \\
    {}  & [$R_{*}$] & [K] & [$\rm{cm}^{-3}$] & {} \\
    \hline
     R Leo & {567.1} & {40} & {$\left[ 1.38\times 10^{3}, 3.88\times 10^{3} \right]$} & $\{0,1,..,5\}$ \\
     {} & {392.9} & {50}    & {$\left[ 2.90\times 10^{3}, 8.11\times 10^{3} \right]$} &  $\{0,1,..,5\}$ \\
     {} & {289.0} & {60}    & {$\left[ 5.39\times 10^{3}, 1.50\times 10^{4} \right]$} &  $\{0,1,..,5\}$ \\
     {} & {123.2} & {100}   & {$\left[ 2.99\times 10^{4}, 8.13\times 10^{4} \right]$} &  $\{0,1,..,5\}$ \\
     {} & {19.9} & {300}    & {$\left[ 1.50\times 10^{6}, 3.19\times 10^{6} \right]$} &  $\{0,1,..,5\}$ \\
    \hline
  \end{tabular}
%  \tablenotemark{a}
  \tablenotetext{a}{The density $n(r)$ is shown as a minimum and maximum range across the set of integer values of $\beta \in \{0,1,..,5\}$ enumerated.}
 \end{table}

\begin{table}[ht!]
  \centering
  \caption{CO emission models for R Leo. \label{tab:CO_RLeo}}
  \begin{tabular}{ccccl}
  \hline
  \hline
   $v_c$ & $v_{\infty}$ & $\beta$ & $\dot{M}$ & {Reference} \\
    {[km/s]}  & {[km/s]} &  & {[$10^{-7} M_{\odot}/{\rm yr}$]} &  \\
    \hline
     3.0  & 8.5 & 5.0 & 1.1 & \citet{success15} \\
      {-} & 5.0 & 0 & 1.8 & \citet{schoier13} \\
      {-} & 6.0 & 0 & 1.0 & \citet{ramstedt14} \\
      {-} & 6.0 & 0 & 2.0 & \citet{gonzalez03} \\
      {-} & 6.0 & 0 & 1.2 & \citet{teyssier06} \\
      {-} & 9.0 & 0 & 0.92 & \citet{debeck10} \\
    \hline
  \end{tabular}
%  \tablenotemark{a}
%  \tablenotetext{a}{Text}
 \end{table}

\begin{table}[ht!]
  \centering
  \caption{Minima and maxima in the radial density profiles $n_k(r)$ across the CO emission models tabulated in Table \ref{tab:CO_RLeo} sampled at or very near the radii listed in Table \ref{tab:t_CSEpara}. \label{tab:CO_RLeo_n}}
  \begin{tabular}{ccccc}
  \hline
  \hline
   Source & r  & T & $n_{\rm min}(r)$ & $n_{\rm max}(r)$  \\
    {}  & [$R_{*}$] & [K] & [$\rm{cm}^{-3}$] & {[$\rm{cm}^{-3}$]} \\
    \hline
     R Leo & {567.1} & {40} & {$1.75\times 10^{3}$} & {$3.82\times 10^{3}$} \\
     {} & {392.9} & {50}    & {$3.72\times 10^3$} & {$7.95\times 10^{3}$} \\
     {} & {289.0} & {60}    & {$6.74\times 10^3$} & {$1.47\times 10^4$}\\
     {} & {123.2} & {100}   & {$3.71\times 10^4$} & {$8.09\times10^4$} \\
     {} & {19.9} & {300}    & {$1.45\times 10^6$} & {$3.11\times 10^6$} \\
    \hline
  \end{tabular}
%  \tablenotemark{a}
%  \tablenotetext{a}{Text}
 \end{table}

\section*{Acknowledgement}
This material is based upon work supported by the National Science Foundation under grant no. NSF-AST 1139950. We thank our colleague, Robert Harris, for his advice and assistance during this project. We thank Roy Lankhaar for collaborative and helpful discussions. This paper makes use of the following ALMA data: ADS/JAO.ALMA\#2011.0.00001.CAL. ALMA is a partnership of ESO (representing its member states), NSF (USA) and NINS (Japan), together with NRC (Canada), MOST and ASIAA (Taiwan), and KASI (Republic of Korea), in cooperation with the Republic of Chile. The Joint ALMA Observatory is operated by ESO, AUI/NRAO and NAOJ. 
The National Radio Astronomy Observatory is a facility of the National Science Foundation operated under cooperative agreement by Associated Universities, Inc. This research made use of Astropy, a community-developed core Python package for Astronomy \citep{AstropyCollab}.This research has also made use of data from the MOJAVE database that is maintained by the MOJAVE team \citep{lister_2018}. This study makes use of 43 GHz VLBA data from the VLBA-BU Blazar Monitoring Program (VLBA-BU-BLAZAR\footnote{
http://www.bu.edu/blazars/VLBAproject.html}), funded by NASA through the Fermi Guest Investigator Program. This paper uses astronomical data from the POLAMI program, which uses the IRAM 30\,m telescope. This work has made use of data from the European Space Agency (ESA) mission
{\it Gaia} (\url{https://www.cosmos.esa.int/gaia}), processed by the {\it Gaia}
Data Processing and Analysis Consortium (DPAC,
\url{https://www.cosmos.esa.int/web/gaia/dpac/consortium}). Funding for the DPAC
has been provided by national institutions, in particular the institutions
participating in the {\it Gaia} Multilateral Agreement.
\software{CADRE \citep{cadre}, MIRIAD \citep{miriad}, SciPy \citep{Scipy}, Astropy \citep{AstropyCollab}, APLpy \citep{APLpy}, Matplotlib \citep{Matplotlib}, NumPy \citep{NumpyGuide,NumpyPaper}}
%\AKnoteM{Do we need to cite numpy and matplotlib? -M}
\newpage
\bibliographystyle{aasjournal}
%\bibliography{sec34,APNsci,Bfield,Line_info,Target_info,CSE_para}
\bibliography{main}
\end{document}